\documentclass[9pt,conference]{IEEEtran}
\IEEEoverridecommandlockouts

\usepackage[bottom=2.1cm, top=1.20cm,left=1.72cm,right=1.72cm]{geometry}
\usepackage{tikz}
\usepackage{contour}

\usepackage{cite}
\usepackage{amsmath,amssymb,amsfonts}
\usepackage{algorithmic}
\usepackage{graphicx}
\usepackage{textcomp}
\usepackage{xcolor}
\usepackage{bbm}
\usepackage{tikz}
\usepackage{multirow}
\usepackage{colortbl}
\usepackage{hhline}
\usepackage{amsmath}
\usepackage{soul}
\usepackage{subfigure}
\usepackage{subcaption}   

\usepackage[table,xcdraw]{}
\usepackage{booktabs} % For formal tables
\usepackage{stfloats}
\usepackage[normalem]{ulem}
\usepackage{enumitem}
\usepackage{flushend}
\usepackage[font=small]{caption}

\usepackage{fancyhdr}
\usepackage[normalem]{ulem}
\usepackage{url}
\usepackage{array}
\usepackage{tikz}
\usepackage{makecell}
\usepackage{booktabs}

\usepackage{caption}
\usepackage{amsthm}
\usepackage{amsmath,amsfonts}
\usepackage{algorithmic}
\usepackage{graphicx}
\usepackage{textcomp}
\usepackage[table,xcdraw]{}
\usepackage{booktabs} % For formal tables
\usepackage{stfloats}
\usepackage[normalem]{ulem}
\usepackage{subfigure}
\usepackage{multirow}
\usepackage{paralist}
\usepackage{balance}
\usepackage{bm}
\usepackage[ruled,linesnumbered]{algorithm2e}
\usepackage{enumitem}
\usepackage{subcaption}
\usepackage{flushend}

\usepackage{url}
\usepackage[hidelinks]{hyperref}
\usepackage{marvosym}

\usepackage[ruled,linesnumbered]{algorithm2e}
\usepackage{cleveref}
\usepackage{colortbl}

\graphicspath{{./graphics/}}

\newcommand{\parlabel}[1]{{\noindent\bf #1}}

\newcommand{\ia}{\mbox{\sc IA}}

\newcommand{\w}{\mbox{\sc W}}

\newrobustcmd*\circled[1]{\tikz[baseline=(char.base)]{
            \node[shape=circle,draw,inner sep=1pt,fill,text=white,minimum size=1.2em] (char) {\textsf{\small #1}};}}

\crefname{figure}{figure}{figures}

\title{NVR: Vector Runahead on NPUs for Sparse Memory Access
\thanks{This paper has been accepted by DAC'2025.} 
\thanks{* \ Equal contribution.}
\thanks{\Letter \ Corresponding author.} 
\vspace{-32pt}
}

\author{
\IEEEauthorblockN{
Hui Wang\textsuperscript{1}\textsuperscript{*},
Zhengpeng Zhao\textsuperscript{4}\textsuperscript{*},
Jing Wang\textsuperscript{1}, 
Yushu Du\textsuperscript{1}, 
Yuan Cheng\textsuperscript{2,3}, 
Bing Guo\textsuperscript{1}, 
He Xiao\textsuperscript{5},\\
Chenhao Ma\textsuperscript{1}, 
Xiaomeng Han\textsuperscript{1}, 
Dean You\textsuperscript{1}, 
Jiapeng Guan\textsuperscript{6}, 
Ran Wei\textsuperscript{6}
Dawei Yang\textsuperscript{2}\textsuperscript{\textrm{\Letter}},
Zhe Jiang\textsuperscript{1}\textsuperscript{\textrm{\Letter}}}

\IEEEauthorblockA{
\textsuperscript{1}\textit{National Center of Technology Innovation for EDA, School of Integrated Circuits, Southeast University}
\textsuperscript{2}\textit{Houmo AI} \\
\textsuperscript{3}\textit{Nanjing University} 
\textsuperscript{4}\textit{Huazhong University of Science and Technology} 
\textsuperscript{5}\textit{Harbin Institute of Technology} 
\textsuperscript{6}\textit{Dalian University of Technology} \\
whmio0115@seu.edu.cn, 
u202114911@hust.edu.cn, 
yuancheng@smail.nju.edu.cn, 
dawei.yang@houmo.ai, 
zhejiang.uk@gmail.com}
\vspace{-28pt}
}

\begin{document}
    
\maketitle

\begin{abstract}
Deep Neural Networks are increasingly leveraging sparsity to reduce the scaling up of model parameter size. However, reducing wall-clock time through sparsity and pruning remains challenging due to irregular memory access patterns, leading to frequent cache misses. In this paper, we present NPU Vector Runahead (NVR), a prefetching mechanism tailored for NPUs to address cache miss problems in sparse DNN workloads. Rather than optimising memory patterns with high overhead and poor portability, NVR adapts runahead execution to the unique architecture of NPUs. NVR provides a general micro-architectural solution for sparse DNN workloads without requiring compiler or algorithmic support, operating as a decoupled, speculative, lightweight hardware sub-thread alongside the NPU, with minimal hardware overhead (under 5\%). NVR achieves an average 90\% reduction in cache misses compared to SOTA prefetching in general-purpose processors, delivering 4x average speedup on sparse workloads versus NPUs without prefetching. Moreover, we investigate the advantages of incorporating a small cache (16KB) into the NPU combined with NVR. Our evaluation shows that expanding this modest cache delivers 5x higher performance benefits than increasing the L2 cache size by the same amount.

% 当计算资源被缩放，内存往往However, these applications typically feature extremely low computation-to-communication ratio and irregular memory accesses, meaning their performance is memory bound, and out-of-order cores provide significant performance advantages over their in-order counterparts.
\end{abstract}

\section{Introduction}
\vspace{-3pt}
Emerging Deep Neural Networks (DNNs), particularly Large Language Models (LLMs), often scale to hundreds of billions of parameters~\cite{1llms, yang2024efficient}, increasingly consuming more storage, memory bandwidth, and computational resources.
Fortunately, these workloads are typically over-parameterised~\cite{han2015learning}, where up to 90\% of parameters in prevalent models can be pruned while maintaining comparable performance~\cite{3sparsity}.
This redundancy presents an opportunity to leverage sparsity to reduce such intensive resource demands. 

\begin{figure}[t]
\vspace{-20pt}
% \centering
\hspace{-10pt}
% \subfigure{\includegraphics[width=0.22\textwidth]{fig1_2.pdf}}
% \subfigure{\includegraphics[width=0.22\textwidth]{fig1_2.pdf}}
\includegraphics[width=\linewidth]{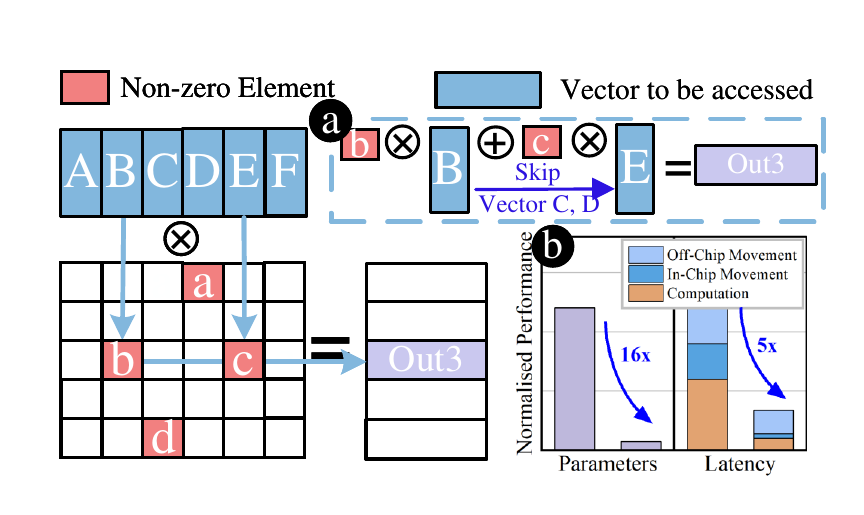}
\vspace{-17pt}
\caption{Sparsity is a widely adopted approach for speedup and energy efficiency through skipping zero in processing, which introduces substantial irregular memory accesses.}
\label{fig:fig1}
\vspace{-17pt}
\end{figure}

Theoretically, more fine-grained sparsity patterns yield higher acceleration by skipping more zero-valued elements.
However, as shown in Fig.~\ref{fig:fig1}~\circled{a}, these patterns introduce unstructured characteristics when directly skipping zero values randomises vector access order, leading to irregular memory access patterns. These irregular patterns frequently result in cache misses, which not only waste main memory bandwidth but also degrade NPU performance.
% \zhenote{Explicitly give numbers regarding the impacts.} 
% As depicted in Fig.\ref{fig:fig1}(b), taking \textcolor{red}{XX} as an example, despite a \textcolor{red}{XX}x reduction in memory access volume, the frequent out-of-bounds SRAM accesses necessitate substantial main memory interactions. 
% This results in the actual memory access benefit being only \textcolor{red}{XX}x, significantly below theoretical expectations. Cache misses impose a barrier to ideal algorithmic performance.
Taking sparse KVCache in LLM workloads~\cite{21doublesparsity} as an example, Fig.~\ref{fig:fig1}~\circled{b} illustrates that despite a 16x parameter reduction, frequent cache misses lead to extensive main memory interactions. 
This results in only a 5x actual speedup, substantially below theoretical expectations. Cache misses thus emerge as a key barrier to sparse workloads.
% 预期多少，然后实际收益只有xxxx

% In light of this, recent work is increasingly exploring the methodology to reduce the side effects of fine-grained sparsity. 

% So recent work are increasingly exploring the implementation and deployment of more fine-grained sparsity, reducing the impact of irregular memory access.\zhenote{Vague}

% \deannote{I fell this version may better.}
% \Guan{hahaha, instead, i think the first one is clearer.}

% \Guan{Perhaps we can use subheadings to make the logic clearer? For example, adding 'Challenges' before the second paragraph and 'Existing works' in the following section.}\mionote{u are right, subheadings are needed.}

% Therefore, there remains significant potential in leveraging sparsity to accelerate execution, provided that the associated cache misses are addressed.
% Recent work is increasingly exploring the methodology to reduce the side effects of fine-grained sparsity. 
% Prior work on hardware-efficient sparse computation implementation has explored various optimization strategies. 
% Therefore, there remains significant potential in leveraging sparsity for acceleration.

\noindent \textbf{Existing Work.}
Recent research has focused on mitigating the side effects of fine-grained sparsity.
Farshchi et al.\cite{farshchi2019integrating} propose a bitmask sparse data format in NVDLA, while %Chen
Eyeriss \cite{chen2016eyeriss} design a run-length encoding for non-zero elements. 
% However, these solutions are often limited to specific hardware or algorithmic implementations. 
Albericio et al.\cite{albericio2016cnvlutin} introduce additional mapping algorithms, regularising memory access patterns through label cache mechanisms. 
Meanwhile, Liu\cite{liu202316} et al. explored sparse domain compute units, developing a butterfly-based architecture for efficient sparse data structure processing.
% \mionote{I think et al. is just for people?}
Whilst most of these approaches attempt to regularise memory access patterns to avoid cache misses, they often introduce substantial area and control logic overhead that outweighs their benefits; therefore, the practical applicability of these algorithm-specific are significantly restricted.

\noindent \textbf{Contributions:} 
Different from all previous works, we tackle the sparsity challenge from a different perspective. 
% We embrace irregular access patterns and leverage runahead prefetching technology directly to mitigate cache misses by pre-executing idle resources on the NPU ahead.
Instead of regularising irregular memory access patterns, our approach embraces the inherent characteristics of sparse computations. We leverage prefetching techniques directly to mitigate the impact of cache misses, rather than attempting to eliminate the irregularity itself, which incurs high overhead and lacks generality. 
% Our key insight is that while irregular memory access patterns are inherent to sparse computations, their performance impact can be minimized through effective prefetching techniques. 
We propose a vector runahead mechanism for NPU (NVR), drawing inspiration from the runahead execution model\cite{8runahead, 8VR}.
% An order-of-magnitude level of cache miss reduction in algorithmic senselessness can be achieved by simply inserting a little prefetcher between the CPU and the NPU.
NVR integrates a lightweight prefetcher decoupled with CPU and NPU, achieving significant cache miss reduction through speculative execution utilising idle NPU units.
%\mionote{if "decoupled" and "utilizing" conflict?}
NVR is a micro-architectural solution requiring no compiler or algorithmic support, remains orthogonal to prior optimisation methods. 
%of optimizing memory access patterns.
% By adopting NVR, we eliminate the need for complex data formats, algorithm modifications, or extensive hardware changes. 
% This approach enables sparse storage, I/O, and computation acceleration without requiring algorithm-specific optimizations.
%This paper presents an approach to directly tackle the fundamental issue of cache misses in sparse computing. 
%Drawing inspiration from the Runahead execution model, we introduce Neural Vector Runahead (NVR). 

% Our contributions can be summarized as follows:
% \vspace{-3pt}
% \begin{itemize}
% \item We conduct comprehensive modelling for sparse DNN workloads and propose NVR, to the best of our knowledge, which is the first system-level prefetching mechanism on NPUs.
% \item We design the architecture of NVR, achieving XXX\% speedup over the SOTA prefetch mechanism on the CPU. 
% \item We explore integrating a tiny cache in NPUs, adding a 16KB cache has the effect of expanding the original 256KB L2 cache by Xx in sparse scenarios. 
% \end{itemize}

To this end, we implement NVR, to the best of our knowledge, the first system-level prefetching mechanism specifically designed for NPUs. 
Our architecture reduces L2 cache misses by 90\% compared to SOTA prefetching techniques in general-purpose processors through the effective prediction of irregular access patterns, with under 5\% area overhead. %, while incurring less than 5\% area overhead.
Furthermore, NVR reduces off-chip memory accesses by 75\% during NPU execution. When combined with NSB (a small cache structure in NPUs), there is an additional 80\% off-chip memory access reduction benefit. 
Our system-level evaluation of LLM workloads further validates the effectiveness of NVR, demonstrating an average 50\% throughput improvement in IO-bound scenarios.

% 只是在decode阶段

% \begin{figure}[t]
% \centering
% \includegraphics[height=0.1\textwidth]{fig2.png}
% \vspace{-5pt}
% \caption{Structured Sparsity and unstructured Sparsity}
% \label{fig:fig2}
% \end{figure}

\begin{figure*}[t]
\vspace{-14pt}
\hspace{-13pt}
% \centering
\includegraphics[width=1.045\linewidth]{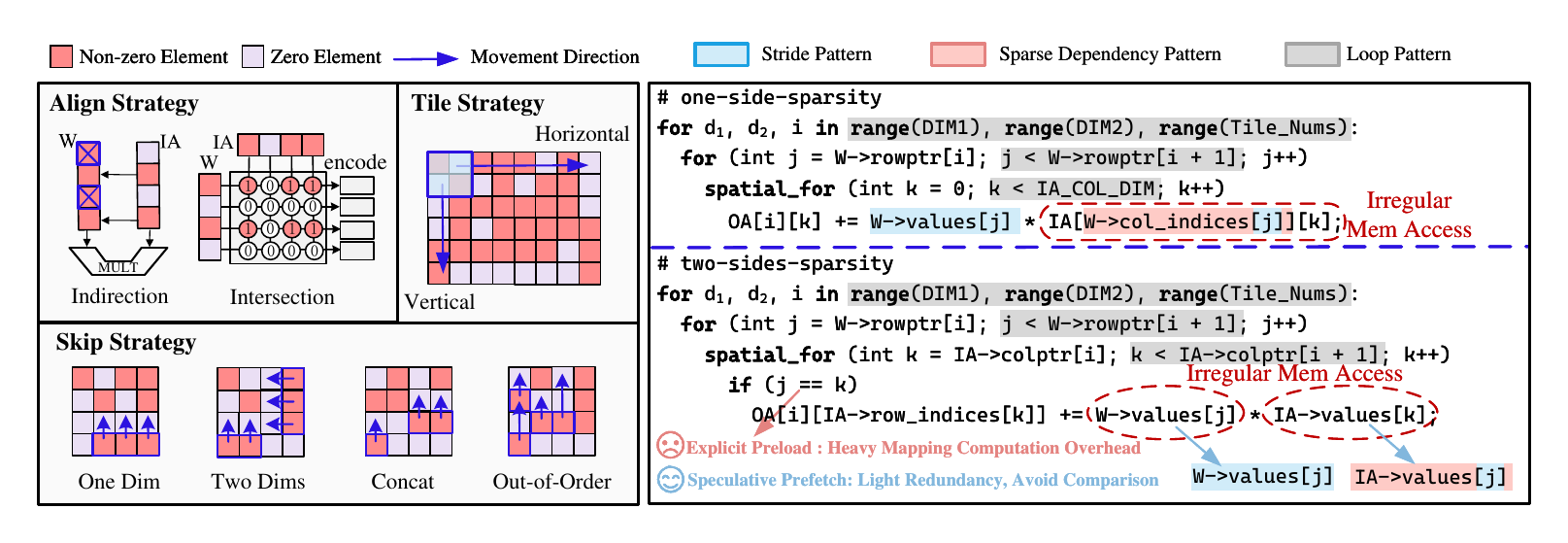}
\vspace{-23pt}
\caption{Sparse Matrix Multiplication can be categorised into one-side-sparsity and two-sides-sparsity patterns, with higher sparsity offering greater speedup potential at the cost of more challenging access patterns. Here, $\texttt{\textbf{spatial\_for}}$ denotes parallel operation on the NPU, while $\texttt{\textbf{IA}}$ (input activation), $\texttt{\textbf{W}}$ (weight), and $\texttt{\textbf{OA}}$ (output activation) represent the input variables, weight parameters, and output results, respectively.}
\vspace{-13pt}
\label{fig:fig2}
\end{figure*}

\par It's worth noting that NVR is not limited to sparse tasks. % While we focus on sparsity as a demonstration, 
% This technique 
It can be applied to any operations exhibiting unstructured characteristics, resulting in irregular memory accesses.
% 如动态的网络计算图

% requires overlapping to increase memory level parallelism (MLP) on NPU.
% This technique can be applied to any operation, effectively overlapping computation with high-latency memory accesses. 

\vspace{-2pt}
\section{Background and Motivation}
\vspace{-2pt}
% \begin{figure}[t]
% \centering
% \includegraphics[height=0.1\textwidth]{fig2.png}
% \vspace{-8pt}
% \caption{Structured Sparsity and unstructured Sparsity}
% \label{fig:fig2}
% \end{figure}

% \begin{figure*}[t]
% \vspace{-20pt}
% \centering
% \includegraphics[width=1\textwidth]{fig3.png}
% \vspace{-20pt}
% \caption{Sparse Matrix Multiplication operations can be categorized into one-side-sparsity and two-sides-sparsity patterns, with higher sparsity offering greater speedup potential at the cost of more challenging access patterns. Here, $\mathrm{\mathbf{spatial\_for}}$ denotes parallel execution across the NPU, while $\mathrm{\mathbf{IA}}$ (input activation), $\mathrm{\mathbf{W}}$ (weight), and $\mathrm{\mathbf{OA}}$ (output activation) represent the input variables, weight parameters, and output results, respectively.}
% \vspace{-10pt}
% \label{fig:fig3}
% \end{figure*}

% \begin{figure}[t]
% \vspace{-20pt}
% \centering
% \includegraphics[width=0.47\textwidth]{fig3_1.png}
% \vspace{-10pt}
% \caption{Structured Sparsity and unstructured Sparsity}
% \label{fig:fig3}
% \vspace{-15pt}
% \end{figure}

% 第一节: sparse 有很多cache miss
% sparse Memory Accesses 为什么会大量cache miss
% 有大量的cache miss, 对于大多数NPU而言对cache miss的容忍性很差。造成的Bottleneck：1. 片外带宽的浪费 2，片内计算单元的空转

% 第二节：对于NPU，prefetch比 CPU更重要
%Another road: replace to explicitly preload and speculatively prefetch 现有方法的不足

% 第三节: prefetch on NPU的挑战（我们面对的挑战）
% 挑战一：workload的多样性，不规则访存的因素多样
% 挑战二：NPU架构的特殊性，
% \subsection{Performance Breakdown: }

Although sparse workloads offer high theoretical memory-level parallelism (MLP), hardware systems struggle to exploit this algorithmic advantage effectively.
As shown in Fig.~\ref{fig:fig1}~\circled{b}, we systematically profile the parameters of Double  Sparsity\cite{21doublesparsity} in LLMs workload and its performance on NPU with a 256KB L2 cache. 
We find that reducing parameters does not lead to proportional decreases in off-chip memory accesses, resulting in out-of-bounds accesses for explicit buffers (like scratchpad) or cache misses for L2 cache in the chip. As these two scenarios are essentially identical, we focus our analysis on cache misses in the following discussion.
% As shown in Fig. \ref{fig:fig2}, we systematically profile the characteristics of sparse workload and their performance on NPU. 
% We find that the challenge of accelerating sparse workloads comes from two challenges: 1) imbalance between compute and memory~2) huge cache miss caused by irregular memory access. 
% Due to reduced data reuse in sparse computations, the decrease in data movement overhead is often proportionally less than the reduction in computational workload, resulting in a shift toward I/O-bound operations.  In this context, the substantial number of cache misses further exacerbates the I/O pressure, becoming a critical performance bottleneck. \mionote{we discuss one challenge in the intro and two here... is this ok?} \mionote{delete the first one}
% 由于稀疏带来的数据复用的减少，所以稀疏带来数据搬运的消耗的缩减比例往往小于计算的缩减比例，造成workload向IO bound偏移。

% Therefore, the memory footprint of features in point cloud networks significantly surpasses CNNs. As shown in Figure \ref{} (right), the memory footprint of the features per point in point cloud networks can achieve up to 16 KB, which is 100× higher than CNNs. Thus the data movement alone can take up over 50\% of total runtime on CPUs and GPUs, as shown in Figure 6 (right).

% These applications typically feature extremely low computation-to-communication ratio and irregular memory accesses, meaning their performance is memory bound, and out-of-order cores provide significant performance advantages over their in-order counterparts.

% \vspace{-8pt}
\vspace{-2pt}
\subsection{Sparse Memory Accesses: Misses Are a Fact of Life}
\label{sc:sc1}
\vspace{-2pt}

In this section, we introduce the sparse computation and analyse the sparse irregular memory access patterns, which lead to a heavy cache miss and NPU stalls.
As an example, Fig.~\ref{fig:fig2}~(right) illustrates a typical sparse matrix multiplication (SpMM), which is common in DNNs, computation with compressed sparse row (CSR) format. The code snippet demonstrates the sparse weight matrix ($\texttt{W}$) selectively indexes input activation ($\texttt{IA}$) vectors for multiply-accumulate operations.
%  exhibits irregular and hard-to-predict memory access behaviours. 
% In this section, we explore the sources of these irregular accesses. % and our approach to effective prediction. 
This workload exhibits several critical factors contributing to cache misses:

\noindent\textbf{Indirect Memory Accesses} Since only non-zero data is computed, the data must be aligned, which involves numerous index dependency chains.
While $\texttt{W[j]}$ accesses are often sequential and can be captured by streaming prefetch, $\texttt{IA[sparse\_func(W[i])]}$ accesses target non-sequential locations. Given that, $\texttt{IA}$ typically spans a large index space that exceeds L2 cache capacity, resulting in frequent cache misses. This alignment pattern is common in across sparsity applications, like hash-table indexing 
%in sparse attention 
and sampling operation in point cloud networks, where $\texttt{sparse\_func}$ is replaced by complex indexing schemes, often requiring dedicated processing units.

\noindent\textbf{Dynamic Loop Boundaries} %Tiling strategy introduces dynamic loop boundaries: 
The tiling strategy refers to reordering element arrangements for hardware computation units (e.g., systolic arrays, vector units). This manifests as irregular loop boundaries. In large dynamic routing architectures such as Mixture-of-Experts (MoE), the memory span between \texttt{rowptr[i]} and $\texttt{rowptr[i+1]}$ can be substantial, posing significant challenges for cache capacity. % For large matrices or high sparsity scenarios, 

% output reorder
\noindent\textbf{Data Shuffle} 
As shown in Fig.~\ref{fig:fig2}~(left), NPUs employ various skip strategies to accelerate computation by bypassing zero elements. 
NPUs with stronger reordering capabilities can group farther distant data elements for unified processing. 
This leads to densely packed, long-stride access patterns within short time intervals, causing catastrophic cache behaviour.
This issue is particularly evident in LLMs' KVCache~\cite{21doublesparsity}, where sparse attention employs TopK selection to retain only the k highest-scoring vectors for computation. For LLMs with extended contexts, KVCache indexing can span several gigabytes, significantly exceeding cache capacity.

% In addition to the aforementioned cache miss sources from computational mode, the data intrinsic sparsity
% Skipping Strategy introduces long-distance sparse skipping patterns: This pattern is prevalent across diverse applications, such as KVCache\cite{} parameter sparsity in LLMs and point cloud network dataset sparsity, where distant memory regions are frequently accessed. In the context of large language models with extended contexts, indexing within the entire KV Cache space potentially spans several gigabytes, far exceeding input buffer capacity. 

% skip 决定了跳过的维度的访存不连续程度，越越不连续
% align 决定了数据转换与Reorder的开销，使得预取要做出一定的妥协
% tile 决定了分界的跳跃程度，越越跳跃

% preload 的预取方式代价太大，所以选择prefetch的方式
% stride pattern: 可以通过
% sparse dependency pattern:
% loop pattern: 因为NPU的黑盒结构所以导致不好检测了

\vspace{-5pt}
% \vspace{-3pt}
% \subsection{More Pressing IO Miss Bound on NPU}
\subsection{Cache-miss Vulnerable NPUs}%Suffering % Sensitivity % Pressing
%Crystal Ball
\vspace{-3pt}

Facing numerous cache misses, NPUs suffer even more severe performance degradation than CPUs and GPUs.
While CPUs leverage Out-of-Order (OoO) execution and reorder buffers (ROB) to tolerate this through fine-grained instruction-level overlap, and GPUs utilise thread-level scheduling for cache misses hiding, NPUs remain limited in such capabilities.
Built primarily on Single Instruction Multiple Data (SIMD) architectures, NPUs possess only coarse-grained instruction parallelism, with compute and memory movement channels decoupled.
As demonstrated in Fig.~\ref{fig:fig5} in the experiment, even with ideal OoO execution on NPUs, it also shows suboptimal performance in handling cache misses.
Critically, the data-parallel nature of NPUs means a cache miss in any vector element stalls the entire processing pipeline, leading to performance degradation on sparse workloads.
% \zhenote{Any quantitive numbers?}

% \noindent\textbf{Explicit Preload and Speculative Prefetch}
NPU memory access overlap is constrained by explicit preload mechanisms and structured scratchpad usage. The preloading process, which requires executing complete load instructions, involves extensive computations and full dependency chains, making it difficult to effectively hide memory latency, particularly in IO-bound scenarios.
While Han et al.\cite{lin2021pointacc} proposed converting explicit memory into a cache, our experiments find this approach insufficient. 
In contrast, NVR’s speculative execution allows flexible prefetching without the need for precise dependency calculations, reducing latency.
% NPU memory access overlap is constrained to explicit preload mechanisms due to their highly structured architecture. They often utilize explicitly managed scratchpad memory for storage and perform row-based computations. 
% While Han et al.\cite{lin2021pointacc} proposes converting explicit memory into a cache to address sparse-induced out-of-memory issues, our experimental evaluation reveals this approach is not insufficient.\mionote{here} 

% Current NPU memory overlap relies on scheduling explicit preload, pre-executing load instructions to move data from main memory to pre-defined buffer positions. 
% However, this approach requires extensive and precise computations, along with complete dependency chains, often making it impractical to overlap memory latency through scheduling alone, particularly in IO-bound scenarios. 
% In contrast, NVR's speculative execution can prefetch without requiring the precise calculations needed to guarantee NPU logic, as long as the prefetch addresses are obtained, offering a more flexible solution.
% \Guan{Add some citations?} \mionote{for what?}
% \mionote{add conventional prefetcher}

% \vspace{-5pt}
\vspace{-5pt}
\subsection{Challenge for NPU Prefetch}
\vspace{-3pt}
% \zhenote{Combine them.}
% 传统预取器有针对Indirect Memory Access, Dynamic Loop Boundaries and Data Shuffle这些的预取，但是直接用到NPU存在多方面的问题.
There are many conventional prefetchers designed for handling irregular memory access patterns mentioned in section \ref{sc:sc1}.
For sparse workloads with abundant irregular memory accesses, simple pattern-based\cite{pattern-based, Feedback-Directed}
%Best-offset} 
or history-based\cite{history, SMS, VLDP} prefetchers often fail to effectively capture and predict these accesses, whereas runahead-based\cite{8runahead, 8CRE, 8PRE} prefetching has been widely adopted as the solution.
Runahead employs speculative execution to hide cache miss latency by prefetching future memory accesses.
VR\cite{8DVR} and DVR\cite{8VR} explore leveraging vector units to perform runahead in parallel, which is a well-suited approach for NPUs that inherently support vector operation instructions.
% However, applying these techniques directly to NPUs presents several challenges.
However, applying these techniques directly to NPUs presents several challenges.

Despite the success of these prefetching techniques in general-purpose processors, the unique characteristics of NPU architectures and their workloads necessitate a fundamentally different approach.
Diverse sparse workloads on NPUs often require customised hardware implementations, such as hash-table-based methods in point cloud processing.
%, exhibiting distinct patterns of irregular and difficult-to-predict memory accesses.
Meanwhile, sparse workloads commonly leverage specialised sparse data formats, like TACO's \cite{parker:2016:meng-thesis} multi-dimensional encoding and SMASH's \cite{kanellopoulos2019smash} hierarchical bitmap-based decompression, handled by dedicated processing units.
This diversity of dependency chains, coupled with the need for parallel execution, poses significant challenges for prefetchers in both pattern capture and overhead. 
A generalised prefetching approach decoupled from specific sparse computation patterns is necessary.
Moreover, NPUs operate with coarse-grained instructions that process entire vectors or matrices. The decoding of such coarse-grained instructions implies weaker locality awareness and significant computational overhead, making traditional instruction-level optimisations less applicable.
\section{NVR: Design Philosophy}
% \vspace{-3pt}

\begin{figure*}[t]
\vspace{-18pt}
\hspace{-15pt}
\includegraphics[width=1.04\textwidth]{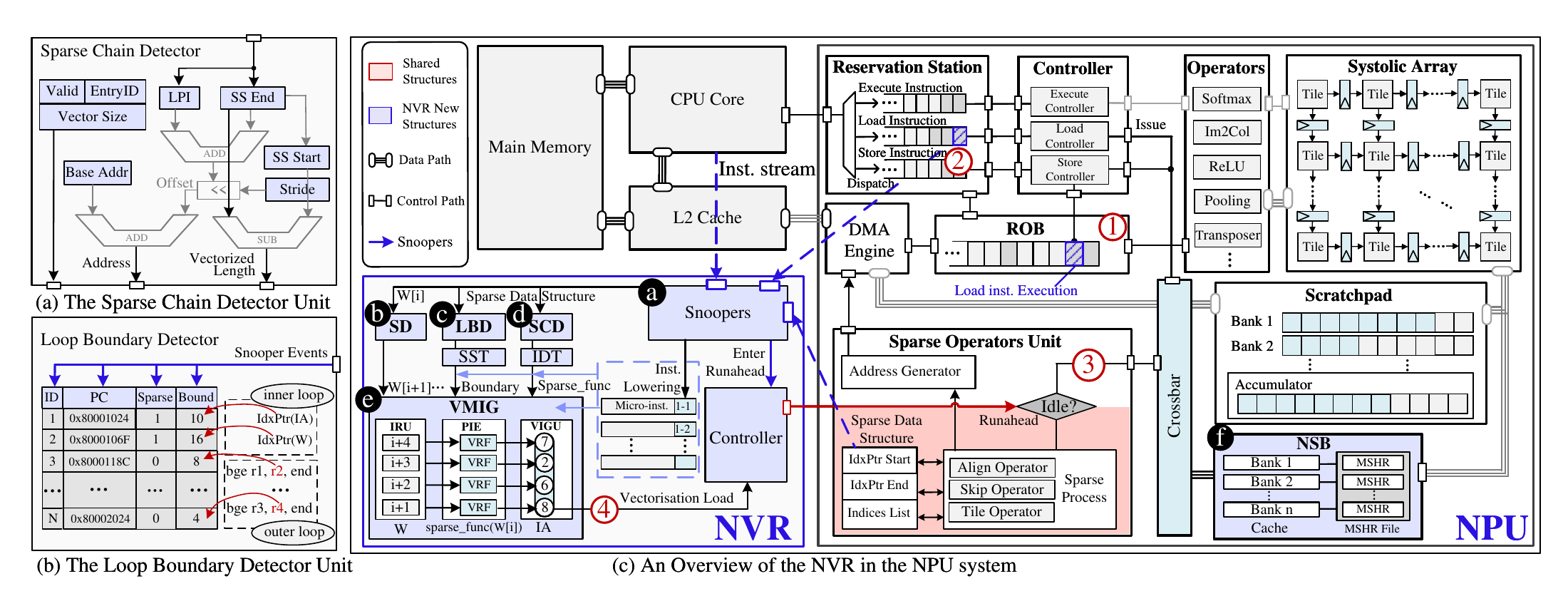}
\vspace{-25pt}
    \caption{NVR micro-architecture and components. Purple blocks represent NVR additions to the system. Red blocks indicate shared components between NVR and NPU, assisting speculative execution during NPU sparse unit idle periods. \circled{b} SD: Stride Detector; \circled{c} LBD: Loop Bound Detector; \circled{d} SCD: Sparse Chain Detector; \circled{e} VMIG: Vectorisation Micro-Instruction Generator; \circled{f} NSB: Non-blocking Speculative Buffer.} %decode sparsity data structure.}
\vspace{-15pt}
\label{fig:fig3}
\end{figure*}

With these challenges in mind, we propose an adaptation of runahead tailored for NPU architectures, addressing three key aspects:

% \noindent \textbf{\underline{Decoupled Philosophy}}:
% 非侵入式原则
\noindent \textbf{\underline{Decoupled and Non-Invasive Philosophy}}:
% Throughput is the key performance metric for NPUs. To maximize it, prefetch should avoid introducing extra control logic that could block NPU execution and achieve the highest possible data parallelism\cite{}.
% For this consideration, our design is decoupling from the pipeline in NPU for future memory accesses. This approach enables speculative execution. 
% Our design achieves non-invasive integration by monitoring CPU and NPU states and instructions. 
% This decoupled architecture eliminates additional compatibility overhead and control logic burden, enabling efficient system operation.
Throughput is the primary metric for NPUs. To maximise it, prefetching mechanisms should avoid introducing additional control logic that stalls execution, ensuring maximum data parallelism\cite{hameed2010understanding}. 
% Our design addresses this by decoupling from the NPU pipeline for future memory access prediction, enabling speculative execution. We achieve non-invasive integration through monitoring of CPU and NPU states and instructions. Operating in parallel with the NPU, our system extracts critical information from load instructions and performs fuzzy prediction, initiating memory requests earlier than the NPU's native execution. Throughout this process, all operations except data transfer remain read-only, ensuring zero interference with NPU execution.
Our design decouples from NPU computation logic, allowing simultaneous speculative execution to proactively generate memory requests.
The system maintains non-invasive integration by passive monitoring the state and extracting load instruction information through read-only operations. 
This approach allows early initiation of memory requests without interfering with NPU execution or custom instruction requirements.
% This decoupled architecture eliminates additional compatibility overhead and control logic burden, enabling efficient system operation.

% our design takes a proactive approach to memory access exploitation

\noindent \textbf{\underline{Coverage-Oriented Philosophy}}:
% While CPU prefetching focuses primarily on latency reduction through cache optimization, NPU prefetching addresses bandwidth limitations to maintain continuous data flow for high-throughput computations. 
In NPU vector operations, computation can proceed only when all data in the batch are ready. 
Our experiments shown in Fig.~\ref{fig:fig8} (a) reveal that the overall cache miss rate decreases significantly faster than per-batch cache miss rates. 
While optimising per-batch cache misses is more challenging, it's crucial for substantial performance improvement.
% \mionote{need show?}\zhenote{Show numbers? Or detailed example?} \mionote{add numbers} 
Consequently, our design prioritises complete batch retrieval, accepting some prefetch redundancy as a reasonable trade-off. 
On the other hand, combining this with fuzzy range loading has the added benefit of reducing the overhead of control logic, like branch prediction. 
% \mionote{need how we tradeoff?}
% \zhenote{No, this is philosophy, discuss the trade-offs in the implementation.}
% \Nnote{Should discussions related to implementation be kept to a minimum in the philosophy section?}

\noindent \textbf{\underline{Micro-Instruction-Level Vectorisation Philosophy}}:
% We exploit NPUs' SIMD architecture through vectorized memory access strategies. 
% \zhenote{This sentence is redundant.}
NPUs' SIMD architecture operates with vectorised instructions, which typically could be decomposed into multiple micro-instructions spanning several cycles.
This fine-grained approach enables precise handling of cache misses while detecting stronger memory access patterns in this granularity.
Leveraging NPUs' native support for vector load instructions, we can bundle memory addresses without additional execution units, improving both prediction accuracy and MLP. 
%This architectural alignment ensures efficient bandwidth utilization and maintains continuous data flow for vector computations.

% NPU instructions tend to be coarse-grained and can be disassembled into a composition of multiple micro-instructions and take multiple cycles.
% Our Runahead implementation operates at the microinstruction level for two key advantages: 
% (1) precise handling of cache misses that occur at different points within a high-level instruction;
% (2) enhanced prefetch prediction accuracy due to stronger memory access patterns at the microinstruction granularity. 

% \begin{figure}
%     \centering
%     \includegraphics[width=1\linewidth]{fig4.png}
%     \caption{Caption}
%     \label{fig:fig4}
% \end{figure}

In summary, vector runahead on NPU has several unique features. We can summarise our runahead design into three Q\&A and introduce detailed micro-architecture in the next session:

\noindent \textbf{\texttt{Q\&A1. When}} to enter runahead mode: Enter runahead when a load instruction in the NPU's ROB executes, prefetching for the next load instruction in the reservation station. NVR executes in parallel with the NPU instruction stream, extracting information through snooping to generate vectorised memory requests ahead of NPU execution. By computing approximate boundaries in advance, it enables early issuance of vector load instructions.

\noindent \textbf{\texttt{Q\&A2. Where}} to execute NVR: 
The NVR is integrated between the CPU and NPU, prefetching speculatively. 
To achieve decoupled and non-invasive operation, the prefetching logic is architecturally separated from the NPU pipeline.
% NVR serves as a separate tiny patch positioned between CPU and NPU, prefetching speculatively.

\noindent \textbf{\texttt{Q\&A3. How}} to prefetch: 
NVR leverages idle computational resources in NPU's sparse processing units to perform approximate dependency chain calculations in parallel. Subsequently, NVR generates native NPU vectorised load instructions for the NPU pipeline, prefetching data into the L1/L2 cache hierarchy. This approach efficiently exploits NPU's vector instructions and architectural characteristics for effective prefetching and analysis.

% \vspace{-5pt}
\section{NVR Micro-Architecture}
% \vspace{-3pt}

% \zhenote{Here, add some descriptions to introduce what we have done, especially why we use Gemini.}

We use Gemmini~\cite{gemmini} as our baseline to demonstrate NVR's applicability to general NPUs, as it embodies typical NPU architecture. %\cite{}.
Its coarse-grained design inherently struggles with cache miss handling, making it ideal for demonstrating NVR's design and effectiveness.

% \vspace{-3pt}
% \vspace{-5pt}
\subsection{Overall Architecture}
\vspace{-1.5pt}
% \vspace{-3pt}
% \mionote{====== I will back, start from here =========}

Fig.~\ref{fig:fig3} illustrates NVR's micro-architectural modifications to the baseline Gemmini configuration, which comprises an in-order core and DNN accelerator sharing a unified L2 cache.
The original Gemmini does not support sparse computation, so we have incorporated a basic implementation of the sparse function to demonstrate our prefetcher's design.
% sparse unit的功能简单介绍
As illustrated in Fig.~\ref{fig:fig3}~\circled{b}, the sparse unit is primarily designed to handle alignment, skipping, and tiling of sparse data, three processing techniques discussed in Section \ref{sc:sc1}.

To support NVR, we augment the baseline with the following structures (purple blocks):
A snooper unit monitoring the status of the CPU and the NPU (Fig.~\ref{fig:fig3}~\circled{a}).
A stride detector (SD) that tracks stream memory access patterns (Fig.~\ref{fig:fig3}~\circled{b}).
A loop boundary detector (LBD) that performs predictive analysis of iteration bounds for both unrolled and nested loop structures (Fig.~\ref{fig:fig3}~\circled{c}).
A sparse chain detector (SCD) identifies indirect memory access dependency chains and computes their corresponding memory addresses (Fig.~\ref{fig:fig3}~\circled{d}).
A vectorisation micro-instruction generator (VMIG) that bundles prefetch sequences with related access patterns into vectorised prefetch operations, optimising bandwidth utilisation (Fig.~\ref{fig:fig3}~\circled{e}).
An optional non-blocking speculative buffer (NSB), which serves as a small in-NPU cache equipped with miss status holding registers (MSHRs) (Fig.~\ref{fig:fig3}~\circled{f}).

The steps of the NVR startup and execution process are numbered with red circles in Fig.~\ref{fig:fig3}.
The NVR initiates runahead execution upon detecting new load instructions in execution within the NPU's ROB. 
The NVR's controller monitors CPU and NPU states via snoopers and triggers runahead request for speculative computations during NPU sparse unit idle periods.
% Using snoopers monitoring CPU and NPU states, the controller triggers runahead requests for speculative computations once the NPU sparse unit turns idle. 
%while maintaining non-intrusive monitoring of critical signals. 
% During execution, the stride detector captures patterns at index $i$ to predict subsequent $A[i]$ values, working in conjunction with the SCD which handles indirect access address generation for $B[sparse\_align(A[i])]$ predictions. 
During execution, the stride detector captures patterns at index $i$ to predict subsequent $\texttt{W[i]}$ values. 
The SCD supports this by generating indirect access address generation for $\texttt{IA[sparse\_func(W[i])]}$ predictions.
To optimise prefetch operations, the LBD analyses loop structures to prevent overruns and enables request packing. 
The VMIG reconstructs decomposed micro-instructions by synthesising information from these components and generates new vectorised prefetching instructions inserted into the NPU pipeline for prefetching operations.
As a complementary approach, the NSB offers an optional mechanism to enhance the prefetching strategy, demonstrating a reduction in cache misses experimentally. 
The following sections detail these components' design.

% \vspace{-5pt}
\vspace{-3pt}
\subsection{Stride Detector (SD)}
% \vspace{-2pt}
\vspace{-1.5pt}

The stride detector is a fundamental pattern recognition unit within the NVR design, tasked with predicting the next batch of addresses for $\texttt{W[i]}$. Its operation relies on identifying consistent striding memory access patterns to facilitate efficient stream prefetching.
The stride detector employs a reference prediction mechanism to track the progression of addresses, similar to reference prediction tables in traditional stride prefetchers. It identifies memory patterns by keeping track of the previous address, stride size, and other control parameters. By leveraging these records, the stride detector predicts subsequent addresses accurately, especially when handling repetitive patterns in workloads. 
In neural network workloads, such as travelling $\texttt{W[i]}$, the typical structure results in relatively fewer branch mispredictions, allowing for high prediction accuracy. 
This is further enhanced by continuously monitoring address differences and maintaining confidence metrics for stride patterns. By predicting the next data accesses effectively, the stride detector not only initiates prefetching requests promptly but also ensures minimal computational overhead, contributing to reduced latency in memory accesses.
% The stride detector is the fundamental pattern recognition unit in NVR design and primarily used to predict the next batch of addresses for $\texttt{W[i]}$.
% It continuously predicts the data memory access pattern for step patterns by stream prefetching. 
% For $\texttt{W[i]}$, neural networks tend to have relatively few branch predictions in their workloads, and a high prediction accuracy can be achieved by recording the address differences and the confidence levels.

% \vspace{-5pt}
% \vspace{-1.5pt}
\subsection{Snoopers and Controller}
\vspace{-2pt}
The snoopers are non-invasive probes used to precisely extract the architectural states of both the CPU and NPU. We monitor three critical signal types:
(1) branch instructions from the CPU, which provide LBD with nested loop context information,
(2) custom load-related instructions in the NPU, which are used to determine the optimal timing for runahead mode activation, and
(3) NPU sparse unit registers, which supply the metadata essential for the NVR prediction mechanisms.
The read-only, non-invasive design of the snoopers maintains architectural integrity by preventing any modifications to NPU computational logic and avoiding interference with the interactions between the CPU and NPU.
Upon entering runahead mode, the controller sends speculative execution requests to the sparse unit. 
When runahead mode is triggered, the controller sends speculative execution requests to the sparse unit, enabling it to proceed with predictive computations. The system then monitors the unit availability, using the snooper infrastructure to retrieve the necessary sparse unit data as soon as it becomes available.

\begin{figure}[t]
\vspace{-5pt}
\hspace{-20pt}
\includegraphics[width=0.55\textwidth]{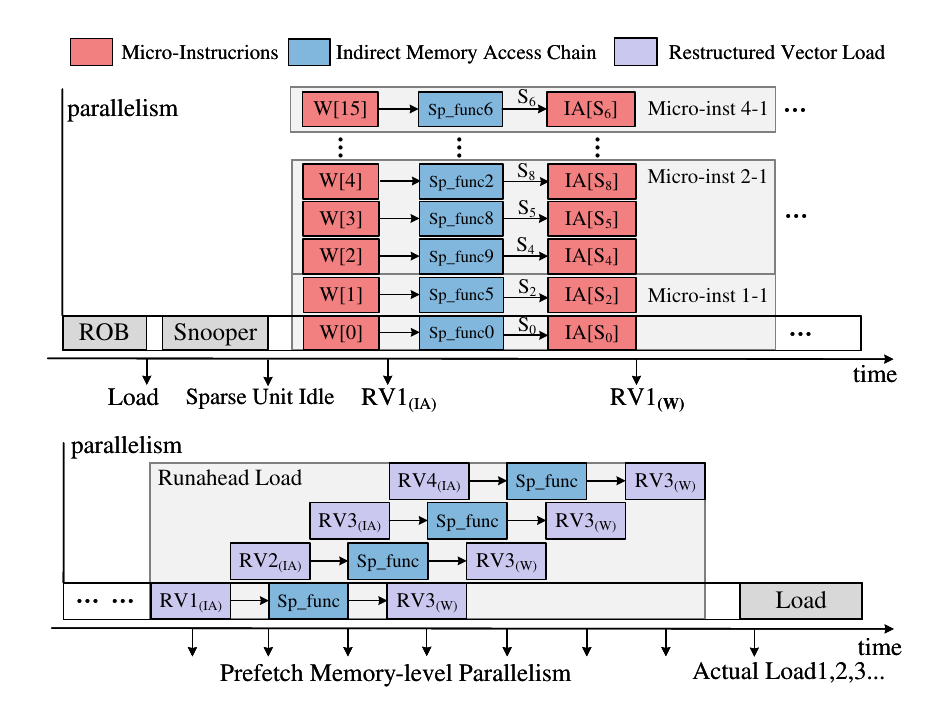}
\vspace{-25pt}
\caption{Vectorisation micro-instruction generation pipeline. Micro-instruction 1-1 represents the first micro-instruction of instruction 1. %(corresponding to a single entry in VMIG). 
Each micro-instruction loads an indeterminate number of data.}
\vspace{-15pt}
\label{fig:fig4}
\end{figure}

% \vspace{-5pt}
% \vspace{-3pt}
\subsection{Sparse Chain Detector (SCD)}
\vspace{-2pt}
% \vspace{-0.5pt}

The SCD identifies and predicts patterns in sparse computations through two critical components: historical information preservation and indirect prefetching pattern learning.
An Indirect Pattern Table (IPT) is maintained to record indirect prefetch patterns, including details such as the Last Prefetch Indirect (LPI) and the sparse structure's start address.
% SCD主要由两部分组成 历史信息的保留 和 间接预取模式的学习
In DNN workloads, computation index patterns typically exhibit locality characteristics. Due to the large volume of data processing, indirect index patterns remain relatively stable over time intervals, often appearing as shallow indirect chains, as formulated below: %\ref{eq:eq1}.
%\mionote{aha?} 
\vspace{-5pt}
\begin{small}
% \begin{equation}
$$
\ia_{\texttt{address}} = \ia_{\texttt{ss\_start}} + (\w_{\texttt{LPI}} << \texttt{stride}) 
\vspace{-5pt}
$$
% \label{eq:eq1}
% \end{equation}
\end{small}
\noindent
, where $\ia_{\texttt{address}}$ represents the predicted indirect access address, $\ia_{\texttt{ss\_start}}$ denotes the base address at the start element of the loop iteration, and $\w_{\texttt{LPI}}$ indicates the value of last prefetched $\w$. Here, $\texttt{stride}$ defines the offset between consecutive memory accesses. 
% For array stride patterns, $\texttt{stride}$ represents the fixed increment of $\w_{\texttt{LPI}}$, typically a multiple of the data format size. 
The prediction for the address of $\ia$ depends on the value of $\w_{\texttt{LPI}}$.

By recording the structural information of ${\texttt{W}}$ and ${\texttt{IA}}$ from previous operations, SCD can effectively track sparse operation chains and predict subsequent indirect prefetching addresses via stride learning. 
This predictive mechanism is particularly beneficial in scenarios with repetitive sparse patterns.
The mechanism requires sparse processing boundaries and current processing indices, such as CSR $\texttt{col\_indices}$ which locates the non-zero elements in columns and $\texttt{rowptr}$ for tracking row start positions from NPU operations, information that is readily available in most sparse data formats.
Unlike traditional prefetchers, which lack access to NPU register-level information and often depend on statistical methods like EWMA\cite{crowder1992ewma}, leading to inefficiencies, our approach uses the NPU's sparse unit to run ahead. This allows us to preserve and leverage historical information, significantly improving the prediction of sparse vector accesses while reducing redundant memory requests.

% SCD maintains historical information through a Sparse Structure Table (SST), where each parallel port has dedicated entry to facilitate concurrent tracking of sparse chains.
% By recording the structural information of ${\texttt{W}}$ and ${\texttt{IA}}$ from previous operations, SCD tracks sparse operation chains and projects subsequent indirect prefetching addresses through stride learning.
% The mechanism requires sparse processing boundaries (e.g., CSR $\texttt{col\_indices}$) and current processing indices (e.g., CSR $\texttt{rowptr}$) from NPU operations, information readily available in most sparse data formats. 
% Traditional prefetchers lack access to NPU register information and depend on historical boundaries or statistical methods like EWMA\cite{crowder1992ewma}, leading to inefficient sparse vector prediction. 
% Our approach overcomes these limitations by running the sparse unit ahead and preserving the historical information. % for IA

% \vspace{-5pt}
\subsection{Loop Bound Detector (LBD)}
\vspace{-2pt}

\begin{figure*}[t]
\vspace{-7pt}
\hspace{-10pt}
% \centering
\includegraphics[width=1.01\textwidth]{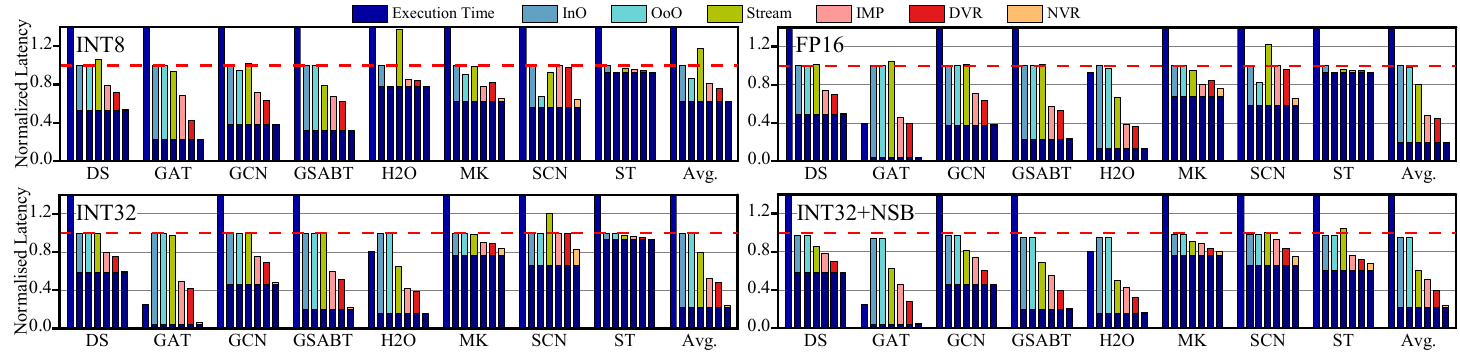}
\vspace{-5pt}
\caption{Normalised wall-clock time latency for each sparse workload. Within each group, each bar from left to right denotes execution in density, in order execution, OoO execution, IMP, DVR, and NVR, respectively. The lower segment indicates the NPU base execution time, whilst the upper segment represents the stall time caused by cache misses.}
\vspace{-15pt}
\label{fig:fig5}
\end{figure*}

DNN workloads inherently consist of multiple nested and unrolled loops, primarily involving matrix and vector operations, where higher-dimensional operations naturally require deeper loop nesting structures. To address the challenges of memory access patterns in these structures, the LBD employs systematic loop behaviour and boundary tracking to optimise prefetch sizing while preventing boundary-crossing invalid prefetches. 

As illustrated in the listing of Fig.~\ref{fig:fig2}, nested loops typically occur during the traversal of rows and columns of a matrix, while unrolled loops are often used in parts of matrix multiplication where multiple indirect chains are executed in parallel. 
As depicted in Fig.~\ref{fig:fig3}~(b), LBD maintains historical information through a Sparse Structure Table (SST), where each parallel port has a dedicated entry to facilitate concurrent tracking of sparse chains. LBD implements a dual-mode boundary prediction, handling both sparse and normal boundaries. The mechanism distinguishes loop hierarchies from inner to outer levels through entry IDs, with each entry maintaining comprehensive loop instruction information, including PC value, boundary values, and operational mode. 
For standard loop boundaries (illustrated in line 1, Fig.~\ref{fig:fig2}~(right)), LBD captures historical boundary information by monitoring register values of jump instructions, as shown with RISC-V B-type branch instructions in Fig.~\ref{fig:fig3}~(b). 
For variable boundaries in sparse computations (shown in line 2, Fig.~\ref{fig:fig2}~(right)), boundary information is dynamically acquired through snoopers from sparse unit registers.
Within this framework, upon detecting loop instructions, the LBD not only uses historical data to predict the boundary but also learns loop boundaries by examining input register values from comparison instructions. 
At the same loop level, instructions are consolidated into memory access requests, while boundary values act as crucial constraints to prevent excessive prefetching. 

% Furthermore, when an address discontinuity is observed in stride detection, the LBD reads the values of registers that will soon be used by comparison instructions. 
% The difference between the observed current values, using the previously identified loop increment, assists in predicting the loop bounds. This prediction assumes that these values were initialized at the start of the loop and have not been modified or spilled before being used in subsequent comparison operations. 
% This proactive prediction during the first iteration of a loop helps ensure more accurate prefetching without waiting for subsequent iterations.
% Within this framework, instructions at the same loop level are efficiently consolidated into several memory access requests, while boundary values serve as critical constraints to prevent excessive prefetching operations.

% \vspace{-5pt}
% \subsection{Vectorization Micro-Instruction Generator (VMIG)}
% \vspace{-3pt}
% % 写细些+加张小图？
% % \mionote{this paragraphs may need more discuss}
% \begin{figure}[t]
% \vspace{-20pt}
% \includegraphics[width=0.475\textwidth]{fig5.png}
% \vspace{-10pt}
% \caption{Vectorization Micro-Instruction Generator}
% \vspace{-5pt}
% \label{fig:fig5}
% \end{figure}

\begin{table}[t]
\vspace{15pt}
\vspace{-5pt}
\Huge
\caption{Hardware Overhead of NVR.} 
\vspace{-5pt}
\centering
\setlength{\arrayrulewidth}{0.75mm} % 设置边框线宽
\renewcommand{\arraystretch}{1.25} % 调整行高
\resizebox{0.48\textwidth}{!}{
\begin{tabular}{|clll|}
\hline
\multicolumn{1}{|c|}{\multirow{3}{*}{\begin{tabular}[c]{@{}c@{}}SD\\ (N=16)\end{tabular}}}   & $48$ bit PC                 & $48N$ bit prev. addr       &    $8N$ bit stride        \\
\multicolumn{1}{|c|}{}                                                                       & $Nlog_2N$ bit entry ID      & $48N$ bit last prefetch addr & $2N$ bit stride conf  \\ \hhline{|~|-|-|-|}
\multicolumn{1}{|c|}{}                                                                       & \multicolumn{3}{c|}{\cellcolor[rgb]{ .906,  .902,  .902}$48 + 16 \times 110$ = 1808 bits}  \\ \hline
\multicolumn{1}{|c|}{\multirow{4}{*}{\begin{tabular}[c]{@{}c@{}}SCD\\ (N=$2\times16$)\end{tabular}}} & $48$ bit PC                 & $48N$ bit ss start   &  $N$ bit valid \\
\multicolumn{1}{|c|}{}                                                                       & $Nlog_2N$ bit entry ID      & $10N$ bit ss offset  & \\ 
\multicolumn{1}{|c|}{}                                                                       & $10N$ bit LPI               & $4N$ bit vector size & \\ \hhline{|~|-|-|-|}
\multicolumn{1}{|c|}{}                                                                       & \multicolumn{3}{c|}{\cellcolor[rgb]{ .906,  .902,  .902} $48+ 32 \times 77 = 2464$ bits}    \\ \hline
\multicolumn{1}{|c|}{\multirow{4}{*}{\begin{tabular}[c]{@{}c@{}}LBD\\ (N=16)\end{tabular}}}  & $48N$ bit PC                & $16N$ bit iteration counter       & $N$ sparse mode   \\
\multicolumn{1}{|c|}{}                                                                       & $Nlog_2N$ bit entry ID     & $16N$ bit increment    & $2N$ level conf  \\
\multicolumn{1}{|c|}{}                                                                       & $16N$ bit loop boundary   & $4N$ bit boundary conf & \\ \hhline{|~|-|-|-|}
\multicolumn{1}{|c|}{}                                                                       & \multicolumn{3}{c|}{\cellcolor[rgb]{ .906,  .902,  .902} $32 \times 1027 = 3424$ bits}    \\ \hline
\multicolumn{1}{|c|}{\multirow{3}{*}{\begin{tabular}[c]{@{}c@{}}VMIG\\ (N=$2\times16$)\end{tabular}}}& $48N$ bit PC                 & $64N$ bit VRF    & $64N$ bit PIE        \\
\multicolumn{1}{|c|}{}                                                                       & $Nlog_2N$ bit entry ID       & $4N+4$ bit IRU   & $256$ bit VIGU         \\ \hhline{|~|-|-|-|}
\multicolumn{1}{|c|}{}                                                                       & \multicolumn{3}{c|}{\cellcolor[rgb]{ .906,  .902,  .902}$260 + 16 \times 184 = 3204$ bits}    \\ \hline
\multicolumn{1}{|c|}{\multirow{3}{*}{\begin{tabular}[c]{@{}c@{}}Snooper\end{tabular}}}& $48$ bit CPU PC  & $64$ bit CPU register       &       \\
\multicolumn{1}{|c|}{}                                                                & $48$ bit NPU PC  & $(48+10+10)N$ sparse structure   &      \\ \hhline{|~|-|-|-|}
\multicolumn{1}{|c|}{}                                                                & \multicolumn{3}{c|}{\cellcolor[rgb]{ .906,  .902,  .902}$160 + 16 \times 68 = 1248 $ bits}   \\ \hline
% \multicolumn{1}{|c|}{\multirow{3}{*}{\begin{tabular}[c]{@{}c@{}}NSB\\ (optional)\end{tabular}}}& 48 bit PC               & 48 bit PC      & 48 bit LP            \\
% \multicolumn{1}{|c|}{}                                                                       & 48 bit PC                 &                & 16 bit LIL                \\ \hhline{|~|-|-|-|}
% \multicolumn{1}{|c|}{}                                                                       & \multicolumn{3}{l|}{\cellcolor[rgb]{ .906,  .902,  .902}xxKiB}                 \\ \hline
\multicolumn{1}{|c|}{Total}                                                                  & \multicolumn{3}{c|}{9.72 KiB + 16 KiB (optional NSB)}                               \\ \hline
\end{tabular}
}
\vspace{-10pt}
\end{table}

% \vspace{-5pt}
\subsection{Vectorisation Micro-Instruction Generator (VMIG)}
\vspace{-2pt}

% Sparse data processing in NPUs compromises SIMD pipeline efficiency, as traditional vector instructions underutilize memory bandwidth due to skipped or zero-valued elements. Employing the original load instructions to prefetch data would result in significant bandwidth wastage due to these sparse data characteristics.
% The VMIG addresses these limitations by reconstructing the micro-instructions within the SIMD load instructions.

In NPU-based sparse data processing, the efficiency of the SIMD pipeline is often compromised because traditional vector instructions fail to fully utilise memory bandwidth when dealing with skipped or zero-valued elements. 
Strictly following NPU runtime loading instructions for prefetching would lead to significant bandwidth wastage. 
Instead, prefetch operations can bypass the rigid loading logic of the NPU, and the VMIG addresses this by restructuring the NPU's native load instructions at the micro-instruction level. 
This approach optimises sparse data patterns, leveraging existing vector load units without additional hardware support. 
% Sparse data processing in NPUs compromises SIMD pipeline efficiency, as traditional vector instructions underutilise memory bandwidth due to skipped or zero-valued elements. Direct adherence to NPU runtime loading instructions for prefetching would result in significant bandwidth wastage. However, prefetch operations need not strictly follow the NPU's rigid \mionote{maybe rigid too fancy?} \Nnote{structured ? may better?} loading logic. Therefore, VMIG restructures the NPU's native load instructions at the micro-instruction level, optimising for sparse data patterns while leveraging existing vector load units without requiring additional hardware support.

% VMIG addresses these limitations through a three-stage pipeline: Micro-instruction Restructuring, Parallel Inference, and Vector Instruction Generation. In the initial restructuring phase, VMIG leverages the LBT while incorporating stride detection for speculative execution. The subsequent inference stage utilizes the IDT from SCD to execute $\texttt{sp\_func}$ predictions concurrently. 
% The hardware implementation comprises three corresponding units: the Instruction Reconstruction Unit (IRU), Parallel Inference Engine (PIE), and Vector Instruction Generation Unit (VIGU). The VIGU synthesizes analyses from both IRU and PIE, dynamically merging four restructured load instructions into a single vector operation using the NPU's native SIMD load capabilities, requiring no additional vector support.

VMIG implements a three-stage pipeline where each stage is executed by dedicated hardware units. 
Initially, the Instruction Reconstruction Unit (IRU) manages the micro-instruction restructuring, using the SST provided by the LBD and integrating stride detection for speculative execution. 
The following stage is handled by the Parallel Inference Engine (PIE), which uses the IPT from the SCD and the Vector Register File (VRF) to predict $\texttt{sp\_func}$ concurrently across multiple data streams. As shown in Fig.~\ref{fig:fig4}, it parallelly executes sixteen consecutive dependency chains that generate $\w$ from $\ia$.
Finally, the Vector Instruction Generation Unit (VIGU) synthesises these restructured load instructions into a single vector operation, dynamically optimising the memory access. 

% VMIG addresses these limitations through a three-stage pipeline architecture, with each stage implemented by dedicated hardware units. The Instruction Reconstruction Unit (IRU) manages the initial Micro-instruction Restructuring phase, leveraging the SST from LBD while incorporating stride detection for speculative execution. 
% The Parallel Inference Engine (PIE) handles the subsequent inference stage, utilising both the IPT from SCD and a vector register file (VRF) to execute $\texttt{sp\_func}$ predictions concurrently across multiple data streams. 
% The Vector Instruction Generation Unit (VIGU) performs the final synthesis by dynamically merging four restructured load instructions into a single vector operation. 
This optimisation strategy leverages the NPU's inherent SIMD load capabilities and vector registers, eliminating the need for additional vector hardware. 
As shown in Fig.~\ref{fig:fig4}, the process is fully pipelined, enhancing MLP through rapid consecutive memory request issuance.
Compared to executing single instructions, this pipelined approach significantly reduces memory access latency and improves bandwidth utilisation. 
The efficiency also depends on the MSHR, which prevents cache miss events from blocking subsequent prefetch operations.

\vspace{-3pt}
\subsection{Non-blocking Speculative Buffer (NSB)}
\vspace{-2pt}
% 加l1cache的理由
For discrete data structures, utilising scratchpad memory incurs substantial logic overhead for data transformation and retrieval operations. However, the characteristics of sparse data patterns present unique opportunities for cache-based optimisation. By strategically storing sparse discrete data in the cache while maintaining continuous data in scratchpad memory (e.g., dense vectors in one-side-sparsity), we can leverage implicit cache line reuse patterns. This cache-based approach naturally accommodates irregular reuse patterns without the costly pre-computation overhead required by scratchpad implementations.
To exploit these characteristics, we introduce NSB, a compact non-blocking cache architecture optimised for discrete element management. Sparse workloads exhibit irregular memory access patterns with extensive index spaces at high sparsity, direct-mapped or low-associativity configurations frequently encounter conflict misses. Thus, we implement a high-way set-associative mapping strategy.
While NSB cannot mitigate L2 cache misses during prefetch operations (as data inherently resides off-chip), it significantly reduces NPU-to-L2 latency and off-chip memory accesses during actual load instruction execution. Experimental results demonstrate that NSB further reduces bandwidth requirements by 5× compared to baseline NVR prefetching.
The design incorporates an MSHR file to manage concurrent memory operations, enabling tracking of outstanding load buffer requests and cache misses. This MSHR infrastructure coalesces multiple outstanding requests to the same cache line, eliminating redundant memory accesses and optimising bandwidth utilisation.

% The design purpose of the non-blocking speculative load buffer is based on the following insights, without the buffer, prefetching competes with computation units for the limited bandwidth of the shared cache, which adversely affects the coverage of prefetching. 
% In contrast, the prefetcher can issue requests to next-level memory through the buffer, while computation units simultaneously access data from the buffer, requests to the shared cache are increased in the event of misses during demand accesses, allowing the majority of memory traffic to be utilized for prefetching. The non-blocking design ensures that regular accesses and prefetching can occur concurrently.

% The Load Buffer is similar to a cache in the CPU but employs a fully associative mapping. Each entry consists of a valid bit, a tag, and data. 
% Miss-status holding registers (MSHRs) are utilized to manage access requests to the Load Buffer. The total size of the Load Buffer is 16 KB.

% \begin{figure}[t]
% \centering
% \includegraphics[width=0.45\textwidth]{fig6.png}
% \caption{VMAG}
% \label{fig:fig5}
% \end{figure}

\vspace{-5pt}
\subsection{Hardware Overhead}
\vspace{-3pt}
Table \ref{tab:tab1} reports NVR's hardware overhead, where $N$ determines the number of parallel entries, matching the vector processing width (default $N$=16). 
The storage overhead is merely 9.72 KiB, negligible relative to the NPU footprint. 
We implement it in RTL and synthesise our design using Synopsys Design Compiler on TSMC 28nm process technology at 2.0 GHz. 
The area overhead is 3\% and 4.6\% relative to the baseline Gemmini architecture, for configurations without and with NSB, respectively. 
These results demonstrate that NVR achieves its performance benefits with minimal hardware cost.

% As $N$, the dominant impactor of MLP and thus performance, grows to 64, the area grows linearly to give XXX KiB total overhead. 
% We evaluate on N = 16 by default but give other values in our evaluation to show the performance-area tradeoff.

% We implemente the NVR's design through RTL code and obtained an area overhead of $0.0305mm^2$, which caused only a 3\% increase in the area compared to Gemmini. 添加NSB的设计则增加4.7\% area. All designs are synthesized using the Synopsys Design Compiler at 2.0 GHz on 28nm TSMC process. 

\section{Evaluation}
% \vspace{-3pt}

\begin{table}[t]
    \centering
    % \vspace{5pt}
    \caption{Sparse Computation Workload.}
    \vspace{-5pt}
    \begin{tabular}{|c|c|c|}
    \hline  
        Workload & Short & Domain  \\
        \hline
        \multicolumn{3}{c}{\vspace{-7pt}} \\
        \hline
        Double Sparsity\cite{21doublesparsity}               & DS     & large language model            \\\hline
        Graph Attention Networks\cite{22GAT}      & GAT    & graph neural networks           \\\hline
        Graph Convolutional Networks\cite{23GCN}  & GCN    & graph neural networks           \\\hline
        Graph Sparse Attention \cite{24GSABT}            & GSABT  & sparse attention                \\\hline
        Heavy-Hitter Oracle\cite{25h2o}           & H2O    & large language model            \\\hline
        MinkowskiNet\cite{26MK}                & MK     & point cloud                     \\\hline
        SparseConvNet \cite{27SCN}                        & SCN    & point cloud                     \\\hline
        Switch Transformer\cite{28ST}             & ST     & mixture of experts              \\\hline
    \end{tabular}
    \vspace{-15pt}
    \label{tab:tab1}
\end{table}

% \begin{figure*}[htbp]
% \centering
% \vspace{-20pt}

% \vspace{-10pt}
% \caption{ Demo }
% \label{fig:fig7}
% \end{figure*}

% \begin{figure}[htbp]
% \centering
% \includegraphics[width=0.5\textwidth]{fig7_2.png}
% \caption{XXXXXXXX}
% \vspace{-20pt}
% \end{figure}

\vspace{-5pt}
\subsection{Experimental Setup}
\vspace{-3pt}
We implement and evaluate NVR by integrating it into a Gemmini-like NPU model constructed using the ScaleSim simulator\cite{scalesim}. 
Additionally, we leverage LLMCompass\cite{llmcompass}, a specialised simulator designed to evaluate hardware optimisations in LLM inference.% workloads.

% \textbf{\textit{Comparison}}
\noindent \textbf{Comparison}
We evaluate our approach against in-order Gemmini (serial execution of load and compute instructions), ideal out-of-order Gemmini (overlapping the load and computation time), and several prefetchers.
We evaluate the following prefetch techniques:
1) stream prefetcher\cite{stream}, the simplest prefetch mechanism based on stride;
2) IMP\cite{imp}, one of the SOTA indirect memory prefetchers, prefetching indirect memory access at the L1 cache;
3) DVR\cite{8DVR}, one of the SOTA runahead techniques, vectorising the same chain of indirect memory accesses across multiple invocations of an inner loop.
4) Our NPU Vector Runahead. 
For a fair comparison, we add the Gemmini custom ISA support for these prefetchers and expand them to the same number of parallels as the NVR. 

% \textbf{\textit{Workload}}
\noindent \textbf{Workload}
Sparsity is widely present across various workloads in DNNs. 
We select several key sources of sparsity, including attention mechanisms\cite{attention, sparseattention}, MoE structures\cite{MoE, glam}, 3D point cloud processing\cite{3Dpoint, dgcnn}, and so on. 
% We collect a variety of models based on their functionalities and extract their linear layers' memory access pattern to serve as representative workloads, as shown in Table \ref{tab:tab1}.
Table \ref{tab:tab1} presents representative workloads extracted from various models' linear layer memory access patterns based on their functionalities and domains.

% \begin{table}
%     \centering
%     \caption{Sparse matrice workload.}
%     \vspace{-5pt}
%     \begin{tabular}{|c|c|c|c|c|}
%     \hline  
%         Workload & Short & Domain & Linear Prop. & Density \\
%         \hline
%         \multicolumn{5}{c}{\vspace{-7pt}} \\
%         \hline
%         Double Sparsity\cite{}& SP &  LLM &         & \\\hline
%                               &     & LLM &         & \\\hline
%                               &     & MoE &         & \\\hline
%                               &     & MoE &         & \\\hline
%                               &     & Sparse Attn. &         & \\\hline
%                               &     & Point Cloud &         & \\\hline
%                               &     & Point Cloud &         & \\\hline
%          LightGaussian\cite{} & LG  & 3DGS &         & \\\hline
%                               &     & GNN  &         & \\\hline
%                               &     & GNN &  & \\\hline
%     \end{tabular}
%     \vspace{-15pt}
%     \label{tab:tab1}
% \end{table}

% \textbf{\textit{Parameter Selection}}

\begin{figure*}[t]
\centering
\vspace{-5pt}
\hspace{-10pt}
\subfigure{\includegraphics[height=2.7cm]{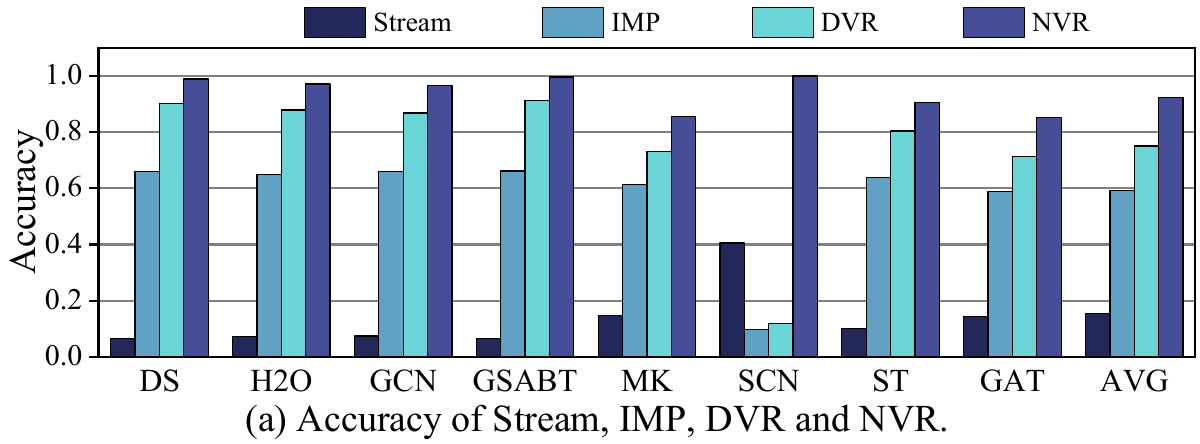}}
\hspace{-7pt}
\subfigure{\includegraphics[height=2.7cm]{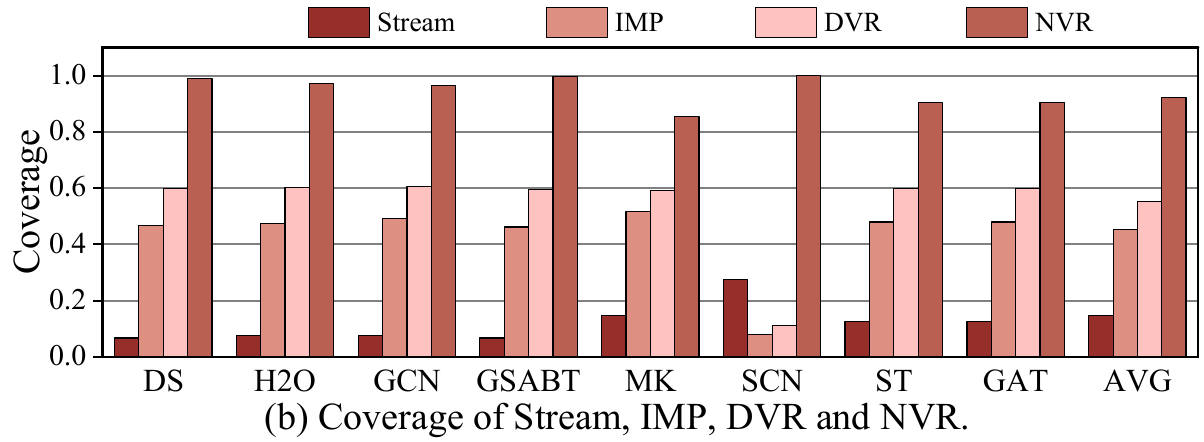}}
\subfigure{\includegraphics[height=2.7cm]{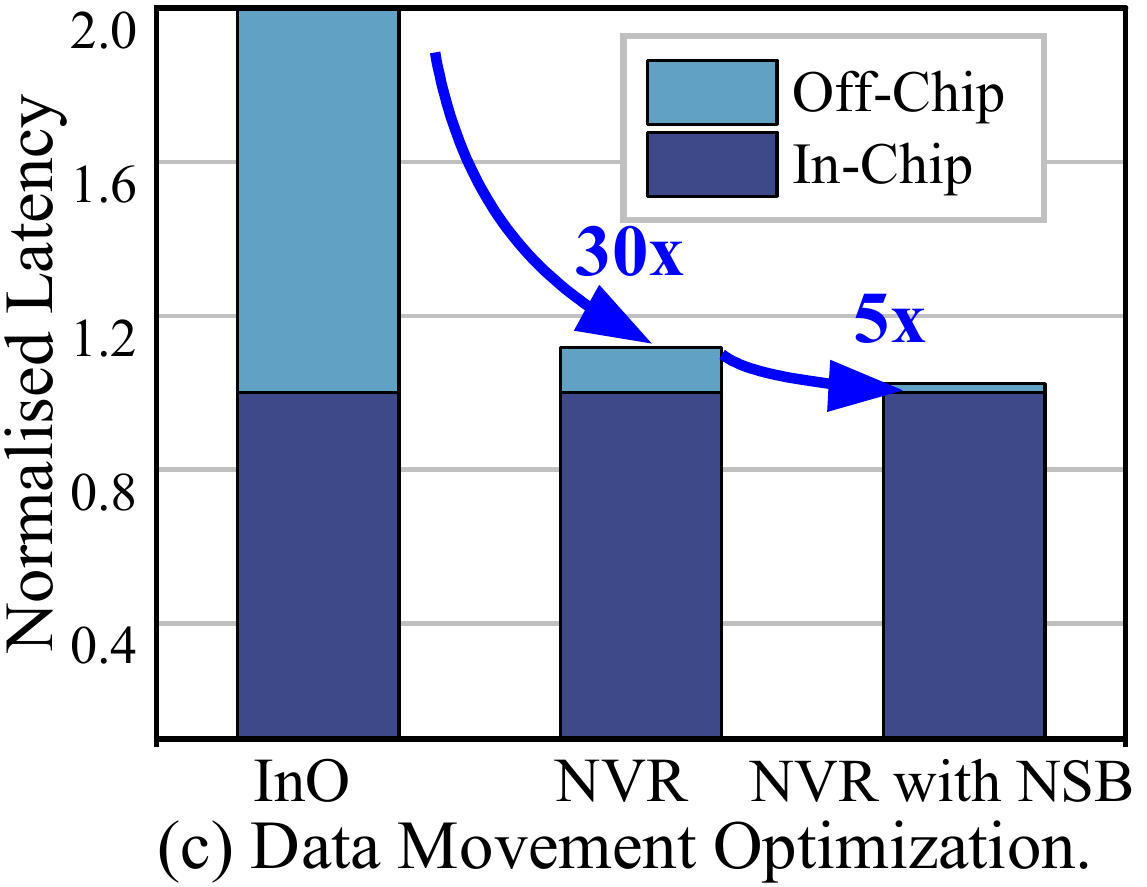}}
\vspace{-8pt}
\caption{Prefetcher accuracy and coverage evaluation and the data movement optimisation results of NVR.}
\vspace{-18pt}
\label{fig:fig6}
\end{figure*}

\begin{figure}[t]
\centering
\vspace{-3pt}
\hspace{-10pt}
\subfigure{\includegraphics[height=2.5cm]{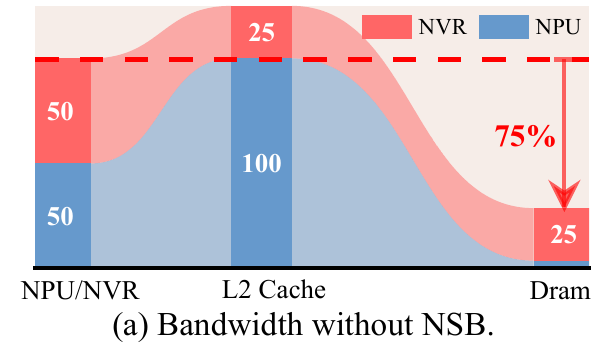}}
% \hspace{-5pt}
\hspace{-10pt}
\subfigure{\includegraphics[height=2.5cm]{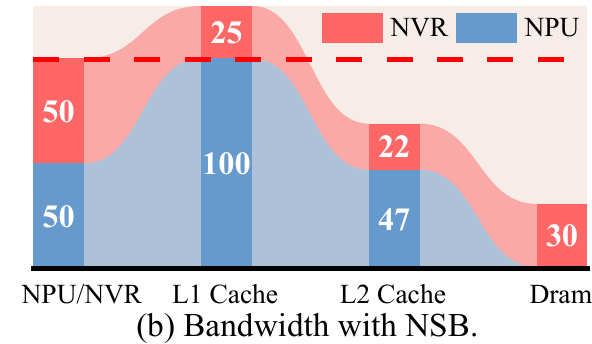}}
\vspace{-8pt}
\caption{Normalised bandwidth allocation without/with NSB.}
\vspace{-5pt}
\label{fig:fig7}
\end{figure}

\begin{figure}[htbp]
\centering
\vspace{-10pt}
% \subfigure[XXXXX]{\includegraphics[height=2.5cm]{fig7_2.png}}
\hspace{-5pt}
\subfigure{\includegraphics[height=3.9cm]{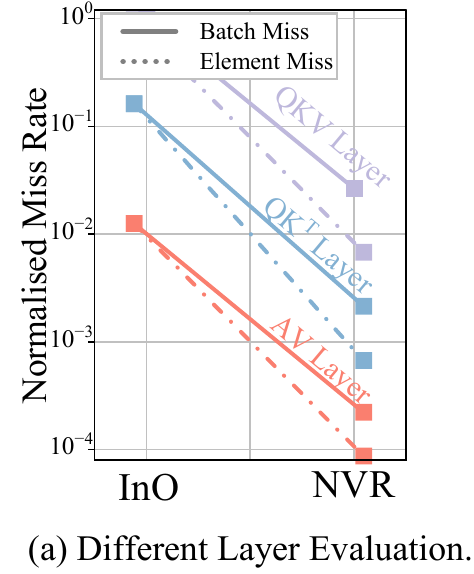}}
\hspace{-17pt}
\subfigure{\includegraphics[height=3.9cm]{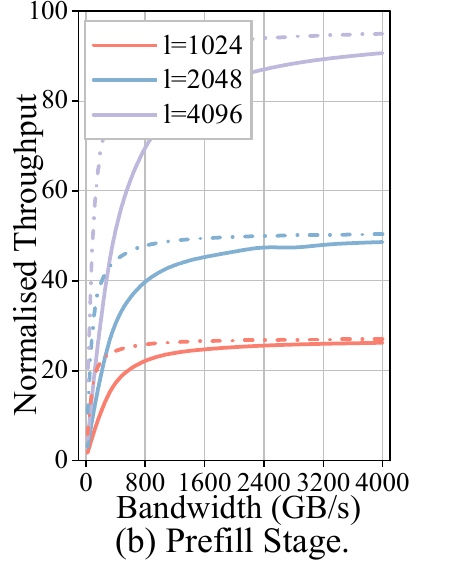}}
\hspace{-20pt}
\subfigure{\includegraphics[height=3.9cm]{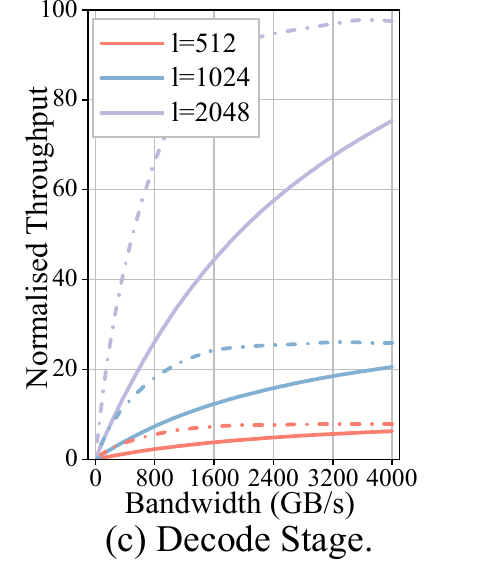}}
% \vspace{-10pt}
% \caption{System-level Evaluation on LLMs \Nnote{height ?}}
% \vspace{-15pt}
% \includegraphics[width=0.48\textwidth]{Fig8.pdf}
% \includegraphics[width=0.48\textwidth,trim={20\% 0 20\% 0},clip]{Fig8.pdf}
\vspace{-10pt}
\caption{System-level Evaluation on LLMs. In (b) and (c), dashed and solid lines represent NVR and baseline performance respectively, where $l$ denotes input/output sequence length.}
\vspace{-15pt}
\label{fig:fig8}
\end{figure}

\vspace{-3pt}
\subsection{Performance Breakdown}
\vspace{-3pt}
Since DNNs commonly employ low bit-width computations, we evaluate different bit-width configurations (INT8, FP16, INT32) to analyse their impact on cache behaviour and system performance, as shown in Figs.~\ref{fig:fig5} (a), (b), and (c). 
Higher data bit-width inherently requires longer
data movement time and larger cache block capacities, increasing the probability of cache misses.

% Due to the widespread use of reduced bit-width computations in DNNs, we evaluate multiple bit-width configurations (INT8, FP16, and INT32) to analyse their impact on cache behaviour and overall system performance, as illustrated in Fig.\ref{fig:fig6} (a), (b), and (c). 
% Higher data bit-width requires longer data movement time and larger cache block capacities, increasing the probability of cache misses. 

Overall, cache miss stalls represent a substantial portion of wall-clock time across most sparse workloads. 
The Switch Transformer presents a notable exception, showing lower cache miss ratios due to its relatively fixed network architecture and block-like data distribution patterns.
For sparse workloads, the NPU's OoO (out-of-order) execution capabilities struggle to effectively overlap data movement with computation to mitigate cache miss latency, as several workloads are inherently IO-bound. 
While prefetching mechanisms generally reduce cache miss stall time effectively, stream prefetchers occasionally introduce performance penalties due to their lower accuracy. 
Notably, our NVR implementation achieve reductions in latency, averaging 98.3\%, 99.2\%, and 97.3\% for INT8, FP16, and INT32, respectively.

As shown in Fig.~\ref{fig:fig5}~(a),  we evaluate INT32 latency variations with NSB enabled. we evaluate latency variations with NSB using INT32 configurations. The red line represents the baseline latency without NSB. 
With NSB enabled, L2 cache miss latency decreased by about 15\%. 
When combined with NVR, NSB achieves a 40\% reduction in cache miss stalls, but performs poorly with conventional stream prefetchers. Thus, NSB activation depends on prefetcher accuracy.

% As illustrated in Figure 6 (d), we perform a detailed evaluation of latency variations with NSB implementation using INT32 configurations. The red line indicates the baseline latency without NSB implementation. Our results demonstrate that with NSB enabled, the L2 cache miss latency decreased by approximately 15\% in most cases for INT32 operations.
% Notably, we observe that the effectiveness of NSB is closely correlated with prefetcher accuracy. When paired with our NVR prefetcher, NSB achieve an average 40\% reduction in cache miss stall time. However, when combine with conventional stream prefetchers, performance is often degraded. Based on these findings, we implement NSB as an optional feature that is conditionally enabled based on prefetcher accuracy thresholds.

\vspace{-5pt}
% \vspace{-3pt}
\subsection{Prefetching Effectiveness and Off-Chip Bandwidth Reduction}
\vspace{-3pt}

This experiment evaluates prefetcher performance across sparse workloads using two key metrics: accuracy and coverage. NVR demonstrates consistently high performance, maintaining both accuracy and coverage rates above 90\% across most workloads. As shown in Fig.~\ref{fig:fig6}~(b), coverage, which is more crucial for NPU performance, presents a greater challenge than accuracy. Benefiting from our fuzzy prefetch strategy, NVR also achieves significantly higher performance in this metric compared to other prefetchers.

The combined effect of accuracy and coverage directly influences prefetch bandwidth and the volume of off-chip memory accesses caused by cache misses during NPU execution.
Figs.~\ref{fig:fig7} (a) and (b) demonstrate the total bandwidth, including prefetch bandwidth. In both cases, compared to the InO baseline, the off-chip memory bandwidth is reduced by around 75\%.
Since the prefetch bandwidth is often more easily overlapped, Fig.~\ref{fig:fig6}~(c) shows the NPU actual load execution time with the memory accesses bandwidth removed. NVR can effectively reduce the off-chip memory accesses by 30x, and with NSB, it can be further reduced by 5x.

\subsection{LLM Inference Evaluation}
\vspace{-3pt}

To validate our system-level optimisations, we evaluate the impact of NVR on the performance of LLMs. The experiments are conducted using the LLMCompass simulator.
As shown in Fig.~\ref{fig:fig8}~(a), the NVR implementation reduces overall latency by decreasing cache miss stall time in the LLMs.
Under NVR, both overall and per-batch cache miss rates decrease exponentially, with the latter showing a slower decay rate.
LLMs are typically composed of two main stages: prefill and decode. It has been proven that the prefill stage is compute-bound, while the decode stage is IO-bound.
Figs.~\ref{fig:fig8}~(b) and (c) demonstrate the benefits of applying NVR to the entire LLM network. 

It is evident that for the compute-bound prefill stage, NVR can expedite reaching maximum throughput, particularly in low-bandwidth scenarios.
For the IO-bound decode stage, the NVR architecture demonstrates an average 50\% throughput improvement through reduced off-chip memory accesses. This enhancement becomes increasingly pronounced with longer output sequences.
% The scatter points represent throughput improvements for different workload configurations before and after optimization. In the IO-bound decoding phase, we achieved XXXx speedup, with benefits becoming more pronounced as the context length increases.

% $$
% I_{\texttt{max}} =\frac{\pi}{\beta}
% $$

\begin{figure}[t]
\vspace{-2pt}
\includegraphics[width=0.48\textwidth,height=0.2\textwidth]{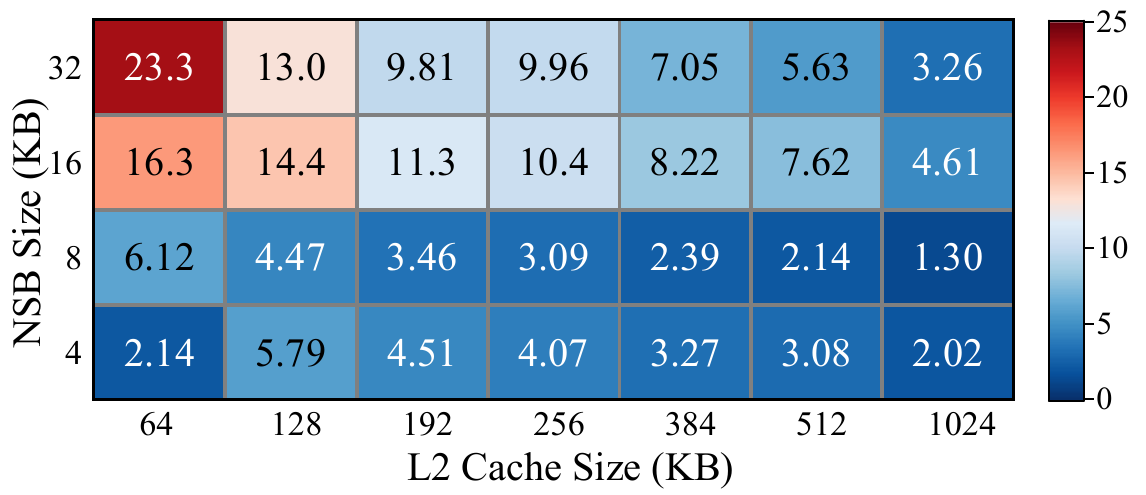}
\vspace{-5pt}
\caption{NSB and L2 cache impact on performance. Performance metric: the inverse of latency and area (higher is better).}
\vspace{-12pt}
\label{fig:fig9}
\end{figure}

% \vspace{-5pt}
\vspace{-3pt}
\subsection{Sensitivity Analyses}
\vspace{-3pt}

To evaluate NSB's efficacy, we conduct comprehensive sensitivity analyses across varying NSB and L2 cache parameters. 
The performance metrics represent the inverse relationship between latency and area, calculated as the product of NSB and L2 Cache dimensions, ensuring equivalent area penalties for both scaling approaches.
As illustrated in Fig.~\ref{fig:fig9}, modest NSB buffer configurations yield substantially higher performance improvements than equivalent L2 cache scaling. Specifically, in a configuration with 256KB L2 cache and 4KB NSB, quadrupling the NSB capacity delivers 5x the performance benefit over scaling the L2 cache to 1024KB. 
% \vspace{-3pt}
\section{Conclusion}
% \vspace{-3pt}

% In this paper, we present NVR, an innovative runahead prefetching mechanism specifically engineered to address the distinctive challenges posed by sparse DNN workloads on NPU architectures.
% NVR employs speculative execution to directly mitigate cache misses, yielding substantial improvements in both performance and efficiency. 
% Our empirical evaluations demonstrate that NVR achieves over 97\% reduction in L2 cache miss latency and delivers up to 4x acceleration in sparse workloads compared NPUs without prefetching capabilities. Furthermore, the integration of NSB amplifies NVR's effectiveness in cache miss reduction. 
% Through vectorised pipelining prefetch operations, NVR significantly reduces off-chip memory accesses, resulting in a 5x enhancement in bandwidth utilisation.

% In this paper, we present NVR, a runahead prefetching mechanism addressing the challenges of sparse DNN workloads on NPU architectures. 
% NVR leverages speculative execution to mitigate the impact of cache misses, improving both performance and efficiency. 
% Our evaluations show that NVR reduces L2 cache miss latency by over 97\% and achieves up to 4x speedup in sparse workloads compared to baseline NPUs. 
% The integration of NSB further enhances NVR's cache miss reduction capability. 
% Through vectorised prefetch operations, NVR reduces off-chip memory accesses, achieving 5x improvement in bandwidth utilisation.

In this paper, we present NVR, a runahead prefetching mechanism addressing the challenges of sparse DNN workloads on NPUs. 
This work demonstrates specialised prefetching techniques for DNN applications on custom architectures beyond general-purpose processors.

\parlabel{Lessons we learnt.}
This work adopted a holistic, workload-driven approach to accelerate the DNNs' executions. 
By workload profiling and architectural analysis, key bottlenecks were identified, guiding targeted system re-architecture. 
Instead of simply adding more hardware resources, this work addressed fundamental inefficiencies, achieving performance gains equivalent to multiple times the benefits of hardware scaling, but with significantly reduced area and power costs. 
The constructed NVR provides key insights that precise, interdisciplinary optimisations rooted in a deep understanding of workload-architecture interactions. 
We believe that such analysis offers valuable insights into bridging architectural and machine learning research, fostering more effective and sustainable design strategies.

\clearpage
% \newpage

\linespread{.98}

\bibliographystyle{IEEEtran}
\bibliography{NVR}

\begin{thebibliography}{10}
\providecommand{\url}[1]{#1}
\csname url@samestyle\endcsname
\providecommand{\newblock}{\relax}
\providecommand{\bibinfo}[2]{#2}
\providecommand{\BIBentrySTDinterwordspacing}{\spaceskip=0pt\relax}
\providecommand{\BIBentryALTinterwordstretchfactor}{4}
\providecommand{\BIBentryALTinterwordspacing}{\spaceskip=\fontdimen2\font plus
\BIBentryALTinterwordstretchfactor\fontdimen3\font minus \fontdimen4\font\relax}
\providecommand{\BIBforeignlanguage}[2]{{%
\expandafter\ifx\csname l@#1\endcsname\relax
\typeout{** WARNING: IEEEtran.bst: No hyphenation pattern has been}%
\typeout{** loaded for the language `#1'. Using the pattern for}%
\typeout{** the default language instead.}%
\else
\language=\csname l@#1\endcsname
\fi
#2}}
\providecommand{\BIBdecl}{\relax}
\BIBdecl

\bibitem{1llms}
Y.~Chang, X.~Wang, J.~Wang, Y.~Wu, L.~Yang, K.~Zhu, H.~Chen, X.~Yi, C.~Wang, Y.~Wang \emph{et~al.}, ``A survey on evaluation of large language models,'' \emph{ACM Transactions on Intelligent Systems and Technology}, vol.~15, no.~3, pp. 1--45, 2024.

\bibitem{yang2024efficient}
D.~Yang, Y.~Wang, R.~Wei, J.~Guan, X.~Huang, W.~Cai, and Z.~Jiang, ``An efficient multi-task learning cnn for driver attention monitoring,'' \emph{Journal of Systems Architecture}, vol. 148, p. 103085, 2024.

\bibitem{han2015learning}
S.~Han, J.~Pool, J.~Tran, and W.~Dally, ``Learning both weights and connections for efficient neural network,'' \emph{Advances in neural information processing systems}, vol.~28, 2015.

\bibitem{3sparsity}
Z.~Yao, S.~Cao, W.~Xiao, C.~Zhang, and L.~Nie, ``Balanced sparsity for efficient dnn inference on gpu,'' in \emph{Proceedings of the AAAI conference on artificial intelligence}, vol.~33, no.~01, 2019, pp. 5676--5683.

\bibitem{21doublesparsity}
S.~Yang, Y.~Sheng, J.~E. Gonzalez, I.~Stoica, and L.~Zheng, ``Post-training sparse attention with double sparsity,'' \emph{arXiv preprint arXiv:2408.07092}, 2024.

\bibitem{farshchi2019integrating}
F.~Farshchi, Q.~Huang, and H.~Yun, ``Integrating nvidia deep learning accelerator (nvdla) with risc-v soc on firesim,'' in \emph{2019 2nd Workshop on Energy Efficient Machine Learning and Cognitive Computing for Embedded Applications (EMC2)}.\hskip 1em plus 0.5em minus 0.4em\relax IEEE, 2019, pp. 21--25.

\bibitem{chen2016eyeriss}
Y.-H. Chen, T.~Krishna, J.~S. Emer, and V.~Sze, ``Eyeriss: An energy-efficient reconfigurable accelerator for deep convolutional neural networks,'' \emph{IEEE journal of solid-state circuits}, vol.~52, no.~1, pp. 127--138, 2016.

\bibitem{albericio2016cnvlutin}
J.~Albericio, P.~Judd, T.~Hetherington, T.~Aamodt, N.~E. Jerger, and A.~Moshovos, ``Cnvlutin: Ineffectual-neuron-free deep neural network computing,'' \emph{ACM SIGARCH Computer Architecture News}, vol.~44, no.~3, pp. 1--13, 2016.

\bibitem{liu202316}
S.~Liu, P.~Li, J.~Zhang, Y.~Wang, H.~Zhu, W.~Jiang, S.~Tang, C.~Chen, Q.~Liu, and M.~Liu, ``16.2 a 28nm 53.8 tops/w 8b sparse transformer accelerator with in-memory butterfly zero skipper for unstructured-pruned nn and cim-based local-attention-reusable engine,'' in \emph{2023 IEEE International Solid-State Circuits Conference (ISSCC)}.\hskip 1em plus 0.5em minus 0.4em\relax IEEE, 2023, pp. 250--252.

\bibitem{8runahead}
O.~Mutlu, J.~Stark, C.~Wilkerson, and Y.~N. Patt, ``Runahead execution: An alternative to very large instruction windows for out-of-order processors,'' in \emph{The Ninth International Symposium on High-Performance Computer Architecture, 2003. HPCA-9 2003. Proceedings.}\hskip 1em plus 0.5em minus 0.4em\relax IEEE, 2003, pp. 129--140.

\bibitem{8VR}
A.~Naithani, S.~Ainsworth, T.~M. Jones, and L.~Eeckhout, ``Vector runahead,'' in \emph{2021 ACM/IEEE 48th Annual International Symposium on Computer Architecture (ISCA)}.\hskip 1em plus 0.5em minus 0.4em\relax IEEE, 2021, pp. 195--208.

\bibitem{lin2021pointacc}
Y.~Lin, Z.~Zhang, H.~Tang, H.~Wang, and S.~Han, ``Pointacc: Efficient point cloud accelerator,'' in \emph{MICRO-54: 54th Annual IEEE/ACM International Symposium on Microarchitecture}, 2021, pp. 449--461.

\bibitem{pattern-based}
N.~Jouppi, ``Improving direct-mapped cache performance by the addition of a small fully-associative cache and prefetch buffers,'' in \emph{[1990] Proceedings. The 17th Annual International Symposium on Computer Architecture}, 1990, pp. 364--373.

\bibitem{Feedback-Directed}
S.~Srinath, O.~Mutlu, H.~Kim, and Y.~N. Patt, ``Feedback directed prefetching: Improving the performance and bandwidth-efficiency of hardware prefetchers,'' in \emph{2007 IEEE 13th International Symposium on High Performance Computer Architecture}, 2007, pp. 63--74.

\bibitem{history}
\BIBentryALTinterwordspacing
Y.~Ishii, M.~Inaba, and K.~Hiraki, ``Access map pattern matching for data cache prefetch,'' in \emph{Proceedings of the 23rd International Conference on Supercomputing}, ser. ICS '09.\hskip 1em plus 0.5em minus 0.4em\relax New York, NY, USA: Association for Computing Machinery, 2009, p. 499–500. [Online]. Available: \url{https://doi.org/10.1145/1542275.1542349}
\BIBentrySTDinterwordspacing

\bibitem{SMS}
S.~Somogyi, T.~Wenisch, A.~Ailamaki, B.~Falsafi, and A.~Moshovos, ``Spatial memory streaming,'' in \emph{33rd International Symposium on Computer Architecture (ISCA'06)}, 2006, pp. 252--263.

\bibitem{VLDP}
M.~Shevgoor, S.~Koladiya, R.~Balasubramonian, C.~Wilkerson, S.~H. Pugsley, and Z.~Chishti, ``Efficiently prefetching complex address patterns,'' in \emph{2015 48th Annual IEEE/ACM International Symposium on Microarchitecture (MICRO)}, 2015, pp. 141--152.

\bibitem{8CRE}
M.~Hashemi, O.~Mutlu, and Y.~N. Patt, ``Continuous runahead: Transparent hardware acceleration for memory intensive workloads,'' in \emph{2016 49th Annual IEEE/ACM International Symposium on Microarchitecture (MICRO)}.\hskip 1em plus 0.5em minus 0.4em\relax IEEE, 2016, pp. 1--12.

\bibitem{8PRE}
A.~Naithani, J.~Feliu, A.~Adileh, and L.~Eeckhout, ``Precise runahead execution,'' in \emph{2020 IEEE International Symposium on High Performance Computer Architecture (HPCA)}.\hskip 1em plus 0.5em minus 0.4em\relax IEEE, 2020, pp. 397--410.

\bibitem{8DVR}
A.~Naithani, J.~Roelandts, S.~Ainsworth, T.~M. Jones, and L.~Eeckhout, ``Decoupled vector runahead,'' in \emph{Proceedings of the 56th Annual IEEE/ACM International Symposium on Microarchitecture}, 2023, pp. 17--31.

\bibitem{parker:2016:meng-thesis}
\BIBentryALTinterwordspacing
P.~A. Tew, ``An investigation of sparse tensor formats for tensor libraries,'' M.Eng. Thesis, Massachusetts Institute of Technology, Cambridge, MA, Jun 2016. [Online]. Available: \url{http://tensor-compiler.org/files/tew-meng-thesis-sparse.pdf}
\BIBentrySTDinterwordspacing

\bibitem{kanellopoulos2019smash}
K.~Kanellopoulos, N.~Vijaykumar, C.~Giannoula, R.~Azizi, S.~Koppula, N.~M. Ghiasi, T.~Shahroodi, J.~G. Luna, and O.~Mutlu, ``Smash: Co-designing software compression and hardware-accelerated indexing for efficient sparse matrix operations,'' in \emph{Proceedings of the 52nd annual IEEE/ACM international symposium on microarchitecture}, 2019, pp. 600--614.

\bibitem{hameed2010understanding}
R.~Hameed, W.~Qadeer, M.~Wachs, O.~Azizi, A.~Solomatnikov, B.~C. Lee, S.~Richardson, C.~Kozyrakis, and M.~Horowitz, ``Understanding sources of inefficiency in general-purpose chips,'' in \emph{Proceedings of the 37th annual international symposium on Computer architecture}, 2010, pp. 37--47.

\bibitem{gemmini}
H.~Genc, S.~Kim, A.~Amid, A.~Haj-Ali, V.~Iyer, P.~Prakash, J.~Zhao, D.~Grubb, H.~Liew, H.~Mao \emph{et~al.}, ``Gemmini: Enabling systematic deep-learning architecture evaluation via full-stack integration,'' in \emph{2021 58th ACM/IEEE Design Automation Conference (DAC)}.\hskip 1em plus 0.5em minus 0.4em\relax IEEE, 2021, pp. 769--774.

\bibitem{crowder1992ewma}
S.~V. Crowder and M.~D. Hamilton, ``An ewma for monitoring a process standard deviation,'' \emph{Journal of Quality Technology}, vol.~24, no.~1, pp. 12--21, 1992.

\bibitem{22GAT}
P.~Veli{\v{c}}kovi{\'c}, G.~Cucurull, A.~Casanova, A.~Romero, P.~Lio, and Y.~Bengio, ``Graph attention networks,'' \emph{arXiv preprint arXiv:1710.10903}, 2017.

\bibitem{23GCN}
T.~N. Kipf and M.~Welling, ``Semi-supervised classification with graph convolutional networks,'' in \emph{International Conference on Learning Representations (ICLR)}, 2017.

\bibitem{24GSABT}
D.~Zhang, J.~Yan, K.~Polat, A.~Alhudhaif, and J.~Li, ``Multimodal joint prediction of traffic spatial-temporal data with graph sparse attention mechanism and bidirectional temporal convolutional network,'' \emph{Advanced Engineering Informatics}, vol.~62, p. 102533, 2024.

\bibitem{25h2o}
Z.~Zhang, Y.~Sheng, T.~Zhou, T.~Chen, L.~Zheng, R.~Cai, Z.~Song, Y.~Tian, C.~R{\'e}, C.~Barrett \emph{et~al.}, ``H2o: Heavy-hitter oracle for efficient generative inference of large language models,'' \emph{Advances in Neural Information Processing Systems}, vol.~36, pp. 34\,661--34\,710, 2023.

\bibitem{26MK}
S.~Brahmbhatt, V.~Mehta, and S.~Ghosh, ``Minkowski sar-unet3d for point cloud semantic segmentation.''

\bibitem{27SCN}
J.~Wang, W.~Li, M.~Zhang, and J.~Chanussot, ``Large kernel sparse convnet weighted by multi-frequency attention for remote sensing scene understanding,'' \emph{IEEE Transactions on Geoscience and Remote Sensing}, vol.~61, pp. 1--12, 2023.

\bibitem{28ST}
W.~Fedus, B.~Zoph, and N.~Shazeer, ``Switch transformers: Scaling to trillion parameter models with simple and efficient sparsity,'' \emph{Journal of Machine Learning Research}, vol.~23, no. 120, pp. 1--39, 2022.

\bibitem{scalesim}
A.~Samajdar, Y.~Zhu, P.~Whatmough, M.~Mattina, and T.~Krishna, ``Scale-sim: Systolic cnn accelerator simulator,'' \emph{arXiv preprint arXiv:1811.02883}, 2018.

\bibitem{llmcompass}
H.~Zhang, A.~Ning, R.~B. Prabhakar, and D.~Wentzlaff, ``Llmcompass: Enabling efficient hardware design for large language model inference,'' in \emph{2024 ACM/IEEE 51st Annual International Symposium on Computer Architecture (ISCA)}.\hskip 1em plus 0.5em minus 0.4em\relax IEEE, 2024, pp. 1080--1096.

\bibitem{stream}
I.~Hur and C.~Lin, ``Memory prefetching using adaptive stream detection,'' in \emph{2006 39th Annual IEEE/ACM International Symposium on Microarchitecture (MICRO'06)}.\hskip 1em plus 0.5em minus 0.4em\relax IEEE, 2006, pp. 397--408.

\bibitem{imp}
X.~Yu, C.~J. Hughes, N.~Satish, and S.~Devadas, ``Imp: Indirect memory prefetcher,'' in \emph{Proceedings of the 48th International Symposium on Microarchitecture}, 2015, pp. 178--190.

\bibitem{attention}
Z.~Niu, G.~Zhong, and H.~Yu, ``A review on the attention mechanism of deep learning,'' \emph{Neurocomputing}, vol. 452, pp. 48--62, 2021.

\bibitem{sparseattention}
A.~Roy, M.~Saffar, A.~Vaswani, and D.~Grangier, ``Efficient content-based sparse attention with routing transformers,'' \emph{Transactions of the Association for Computational Linguistics}, vol.~9, pp. 53--68, 2021.

\bibitem{MoE}
Z.~Chen, Y.~Deng, Y.~Wu, Q.~Gu, and Y.~Li, ``Towards understanding mixture of experts in deep learning,'' \emph{arXiv preprint arXiv:2208.02813}, 2022.

\bibitem{glam}
N.~Du, Y.~Huang, A.~M. Dai, S.~Tong, D.~Lepikhin, Y.~Xu, M.~Krikun, Y.~Zhou, A.~W. Yu, O.~Firat \emph{et~al.}, ``Glam: Efficient scaling of language models with mixture-of-experts,'' in \emph{International Conference on Machine Learning}.\hskip 1em plus 0.5em minus 0.4em\relax PMLR, 2022, pp. 5547--5569.

\bibitem{3Dpoint}
M.~Gadelha, R.~Wang, and S.~Maji, ``Multiresolution tree networks for 3d point cloud processing,'' in \emph{Proceedings of the European Conference on Computer Vision (ECCV)}, 2018, pp. 103--118.

\bibitem{dgcnn}
A.~V. Phan, M.~Le~Nguyen, Y.~L.~H. Nguyen, and L.~T. Bui, ``Dgcnn: A convolutional neural network over large-scale labeled graphs,'' \emph{Neural Networks}, vol. 108, pp. 533--543, 2018.

\end{thebibliography}

\end{document}